\documentclass[a4paper,11pt]{article}

\usepackage[margin=1in]{geometry}

\usepackage[english]{babel}
\usepackage{amssymb}
\usepackage{amsmath}
\usepackage{amsthm}
\usepackage{psfrag}
\usepackage[T1]{fontenc}
\usepackage{ae,aecompl}
\usepackage[colorlinks]{hyperref}
\usepackage{subfigure}
\usepackage{appendix}
\usepackage{natbib}
\hypersetup{colorlinks,breaklinks=true,citecolor=blue,linkcolor=blue}
\usepackage{graphicx}
\usepackage{color}

\newcommand{\Pran}{\textit{Pr}}
\newcommand{\Ek}{\textit{Ek}}
\newcommand{\Ra}{\textit{Ra}}

\newcommand{\Ro}{\mbox{\textit{Ro}}}

\newcommand{\Rey}{\mbox{\textit{Re}}}
\newcommand{\Nu}{\mbox{\textit{Nu}}}

\newcommand{\vel}{\boldsymbol{u}}

\newcommand{\pdt}[1]{\frac{\partial #1}{\partial t}}
\newcommand{\vect}[1]{\boldsymbol{#1}}
\providecommand\bnabla{\boldsymbol{\nabla}}
\providecommand\bcdot{\boldsymbol{\cdot}}

\newcommand\eg{e.g.\ }
\newcommand\ie{i.e.\ }

\title{Jets and large-scale vortices in rotating Rayleigh-B\'enard convection}

\author{C\'eline Guervilly$^1$ \& David W. Hughes$^2$ \vspace{0.2cm}
\\  {\footnotesize $^1$School of Mathematics, Statistics and Physics, Newcastle University, Newcastle Upon Tyne, NE1 7RU, UK}
\\  {\footnotesize $^2$ Department of Applied Mathematics, University of Leeds, Leeds LS2 9JT, UK} }

\begin{document}
\maketitle

\begin{abstract}
	One of the most prominent dynamical features of turbulent, rapidly-rotating convection is the formation of large-scale coherent structures, 
driven by Reynolds stresses resulting from the small-scale convective flows.
	In spherical geometry, such structures consist of intense zonal flows that are invariant along the rotation axis. 
	In planar geometry, long-lived, depth-invariant structures also form at large scales, but, in the absence of horizontal anisotropy, they consist of vortices that grow to the domain size. 
	In this work, through the introduction of horizontal anisotropy 
	into a numerical model of planar rotating convection by the adoption of unequal horizontal box sizes (i.e.\ $L_x \le L_y$, where the $xy$-plane is horizontal), 
	we investigate whether unidirectional flows and large-scale vortices can coexist.
	We find that only a small degree of anisotropy is required to bring about a transition from dynamics dominated by persistent large-scale vortices to dynamics dominated by persistent unidirectional flows parallel to the shortest horizontal direction.
	When the anisotropy is sufficiently large, the unidirectional flow consists of multiple jets, generated on a timescale smaller than a global viscous timescale, thus
	signifying that the upscale energy transfer does not spontaneously feed the largest available mode in the system.
	That said, the multiple jets merge on much longer timescales.
	Large-scale vortices of size comparable with $L_x$ systematically form in the flanks of the jets and can be persistent or intermittent.   
	This indicates that large-scale vortices, either coexisting with jets or not, are a robust dynamical feature of planar rotating convection.

\end{abstract}

\section{Introduction}
In the liquid core of planets, heat is transported primarily by convection. 
Much effort has been dedicated to the study of the properties of convection, in particular  
because of its crucial implications for the thermal evolution of planets and the generation of magnetic fields \citep[\eg][]{Jon07,Aurnou2015}.
Convection is strongly affected by the rapid rotation of the planet via the action of the Coriolis force,
which tends to inhibit small-scale motions along the direction of the rotation axis. Owing to the very low fluid viscosity, flows in planetary 
cores are turbulent, although the nonlinear inertial effects are relatively weak compared with the Coriolis force. 
Rotationally-constrained convection in planetary cores is thus characterised by small Ekman numbers ($\Ek=\nu/2\Omega d^2$, where $\nu$ is the fluid kinematic viscosity,
$\Omega$ the rotation rate and $d$ is the size of the system), large Reynolds numbers ($\Rey=Ud/\nu$, where $U$ is a typical flow velocity at the lengthscale $d$) and
small Rossby numbers ($\Ro=\Ek\Rey=U/2\Omega d$).
Under these conditions, the convective flows take the form of tall narrow columns aligned with the rotation axis.
The concerted action of the Reynolds stresses resulting from the small-scale convective columns drives large-scale coherent structures,
which can become prominent dynamical features.
The occurence of these large-scale structures in experimental and numerical models of turbulent flows \citep[\eg][]{Heimpel05,Read2015} attracts much interest 
because they are observed in many geophysical and astrophysical bodies \citep[\eg][]{Por03,Dyudina08}.
Although the large-scale flows are usually directed perpendicularly to the direction of gravity, they can nonetheless affect the convective heat transport by 
shearing the convective cells or by modifying the local properties of the convection \citep[\eg][]{Guervilly2014,Yad16}. 
Understanding the formation and long term evolution of the large-scale flows is therefore an essential part of the description of a convective system.
Furthermore, large-scale flows can promote the generation of coherent magnetic fields at the system size while the underlying small-scale turbulent flows may, of themselves, be able to generate only fluctuating small-scale magnetic fields \citep{Guervilly2015, Guervilly2017}.

In spherical rotating convection, the large-scale flows consist of intense azimuthal and axisymmetric (\ie zonal) flows that are invariant along the rotation axis \citep[\eg][]{Gil77}. 
The presence of the curved spherical boundaries is crucial for the production of such zonal flows \citep{Bus82}. 
In planar geometry in a horizontally periodic domain, 
long-lived, depth-invariant structures also form at large scales when the horizontal box sizes are equal. 
However, the absence of horizontal anisotropy in this case ensures that no persistent unidirectional flow emerges spontaneously; the large-scale structures therefore consist of vortices that grow to the box size \citep[\eg][]{Chan07}.
These large-scale vortices (LSVs) form whether the rotation axis is aligned or inclined with respect to the direction of gravity \citep{Chan13}.
In this work, we are interested in the possible relation between unidirectional flows and LSVs and whether these two types of large-scale
coherent structures can coexist. 
In simulations of rotating spherical convection, the formation of tall vortices at scales larger than the typical convective size has not been observed. 
In a Cartesian domain, the size of the LSVs is limited only by the horizontal extent of the periodic domain. If the upscale energy transfer were allowed to continue
to larger scales, the LSV would eventually feel either the latitudinal variation of the Coriolis parameter (the $\beta$-effect) or the slope of the vertical boundaries.
The absence of LSVs in spherical models suggests that the large-scale dynamics prefers unidirectional flows rather than vortices when possible.

The convective flows in a turbulent rotating system are three-dimensional (3D), but strongly anisotropic owing to the rapid rotation.
Results from two-dimensional (2D) turbulence might thus be helpful in explaining or predicting some aspects of the dynamics of convective systems 
on timescales longer than the rotation period when $\Ek\ll1$, $\Ro\ll1$.
In particular, the LSVs obtained in 3D convective systems bear a strong resemblance to the condensates that form in 2D turbulence 
from an upscale kinetic energy transfer \citep{Rubio14}. 
However, the analogy with 2D turbulence remains only partial in this particular case because the convective flows that feed energy into the LSVs are anisotropic but certainly not depth-invariant
\citep{Guervilly2014}.
The idea that unidirectional flows are preferred to vortices on the large scales once the horizontal symmetry is broken
is supported by studies of forced 2D turbulence in which the gradual increase of the $\beta$-effect leads first to the weakening of the coherent vortices, 
and eventually to their disappearance, while unidirectional flows emerge \citep{Maltrud1991}. 
However, a counter-example to the weakening of vortices in the presence of symmetry breaking is demonstrated by the recent study of \citet{Frishman2017}, who considered a forced 2D turbulence model with anisotropy introduced by using a rectangular (non-square) periodic domain; with this geometry,
they found that both unidirectional flows and vortices emerge at large scales and coexist on long timescales. 
Here we study whether this remarkable result can carry over to three-dimensional flows.

The two main objectives of the present paper are to determine whether the LSVs that form in rotating planar convection can persist when the symmetry between the horizontal directions is broken and also to determine whether unidirectional flows 
and LSVs can coexist on long timescales. There are a number of ways of breaking the horizontal symmetry. Considering convection with a tilted rotation axis, thereby modelling mid latitudes, is an obvious physically motivated approach. Here though, we choose to introduce a distinction between the horizontal directions in a three-dimensional Cartesian model of rotating Boussinesq convection by considering rectangular, as opposed to square, horizontal domains. Horizontal anisotropy is thus imposed on the system by allowing larger lengthscales to exist in one horizontal direction than the other. This formulation is computationally straightforward to implement, thus allowing a thorough exploration of parameter space, and also allows us to make comparisons with the forced 2D turbulence study of \citet{Frishman2017}.

The layout of the paper is as follows. The mathematical formulation of the problem is briefly described in \S\ref{sec:model}.
The formation of unidirectional flows and LSVs is described in \S\ref{sec:jetLSV}. The box aspect ratio required for the transition
from LSVs to unidirectional flows is quantified in \S\ref{sec:anisotropy} and the selection of the lengthscale of
the large-scale flow is described in \S\ref{sec:merging}. The occurence of bistable states is presented in \S\ref{sec:bistability}.
In \S\ref{sec:domain}, we establish the domain of existence of the large-scale flows. Finally, in \S\ref{sec:heat}, we discuss the effect of the large-scale flows on the 
convective heat transfer. A conclusion is given in \S\ref{sec:ccl}.

\section{Mathematical formulation}
\label{sec:model}

We use a local planar model of rotating Boussinesq convection, as in \citep{Guervilly2014}. The computational domain is three-dimensional and periodic in the horizontal directions. A vertical temperature difference, \mbox{$\Delta T$}, is imposed across the layer of depth $d$. In the horizontal plane, the size of the domain is $L_x d$ in the $x$ direction and $L_y d$ in the $y$ direction. The gravitational field is uniform, \mbox{$\vect{g} = - g \vect{e}_z$}. The rotation vector is \mbox{$\Omega \vect{e}_z$}. The fluid has kinematic viscosity $\nu$, thermal diffusivity $\kappa$,  density $\rho$ and thermal expansion coefficient $\alpha$, all of which are constant. Lengths are scaled with $d$, times with \mbox{$1/(2\Omega)$} and temperature with \mbox{$\Delta T$}. The system of dimensionless governing equations thus becomes
\begin{eqnarray}
	\pdt{\vel} + \vel \bcdot \bnabla \vel + \vect{e}_z \times \vel & = &
	- \bnabla p
	+ \Ek \nabla^2 \vel 
	+ \frac{\Ra \Ek^2}{\Pran} \theta \, \vect{e}_z ,
	\label{eq:u}
	\\
	\nabla \cdot \vel &=& 0 ,
	\\
	\pdt{\theta} + \vel \bcdot  \bnabla \theta - u_z &=& \frac{\Ek}{\Pran} \nabla^2 \theta ,
	\label{eq:theta}
\end{eqnarray}
where \mbox{$\vel=(u_x,u_y,u_z)$} is the velocity field, $p$ the pressure, and $\theta$ the temperature perturbation 
relative to a linear background profile. 
The dimensionless parameters are the Rayleigh number, \mbox{$\Ra = \alpha g \Delta T d^3/\kappa \nu$},
the Ekman number, \mbox{$\Ek = \nu/2\Omega d^2$},
and the Prandtl number, \mbox{$\Pran = \nu/\kappa$}.
The upper and lower boundaries are taken to be perfect thermal conductors,
impermeable and stress free.
Stress-free boundary conditions, where boundary friction is absent, favour the emergence of large-scale flows of large amplitude 
in both planar \citep[][]{Stellmach2014,Kunnen2016} and spherical geometry  \citep[\eg][]{Yad16}.
Equations~(\ref{eq:u}) -- (\ref{eq:theta}) are solved using the pseudospectral code described in detail in \citep{Cattaneo03}. 

\section{Results}

\subsection{Jets and large-scale vortices}
\label{sec:jetLSV}

For most of the results discussed in the paper, the parameters are fixed as $\Ek=10^{-5}$, $\Ra=3\times10^8$, $\Pran=1$, $L_x=1$, with variations only in $L_y$. The Ekman and Rayleigh numbers are varied only in \S\,\ref{sec:merging} and \S\,\ref{sec:bistability}, in order to integrate the system over many viscous timescales, and in \S\,\ref{sec:domain}, in order to determine the domain of existence of the large-scale flows; furthermore, we perform one run with $L_x=2$ and $L_y=4$, which we discuss in the conclusion. The value of the Rayleigh number $\Ra=3\times10^8$ corresponds to $7.4\Ra_c$, where $\Ra_c$ is the critical Rayleigh number at the linear onset of convection \citep{Chandrasekhar61}.

Table~\ref{tab:Ek1e5} contains some global quantities obtained from simulations with $L_y$ in the range $1\leq L_y \leq 8$, together with the numerical resolutions used. The root mean square (r.m.s.)\ amplitude of the velocity is measured by the Reynolds number $\Rey = \langle  |\vel| \rangle^{1/2}\Ek^{-1}$, and
 the r.m.s.\ amplitude of the $i$-component of the 
velocity by $\Rey_i = \langle u_i^2 \rangle^{1/2}\Ek^{-1}$, where the angle brackets denote volume and time averages. 
The inclusion of the Ekman number in these definitions is due to our choice of dimensional units. In order to confirm that the resolution was adequate, and also to verify that the results were not critically dependent on the number of spectral modes adopted, provided this was sufficiently large, we simulated the case of $L_y=3$ with both $512$ and $768$ modes in the $y$ direction. As can be seen from table~\ref{tab:Ek1e5}, the differences between the two simulations are small.

\begin{table}
\small
\begin{center}
\begin{tabular}{c c c c c c c c c c}
\hline \hline
$L_y$ & $\Rey_x$ & $\Rey_y$ &  $\Rey_z$ & $\Rey$ & $\Nu$ & $\Gamma$ & $\alpha$  & $n$ & $N_x\times N_y\times N_z$
\\ \hline 
  $1$ & $2469$ & $2474$ & $910$ & $3612$ & $28.2$ & $5.3$ & $0.00$ & $0$ & $256\times256\times129$ \\
  $1.02$ & $2589$ & $2443$ & $917$ & $3676$ & $28.4$  & $5.4$ & $0.06$ & $0$ & $256\times256\times129$ \\ 
  $1.05$ & $2811$ & $2298$ & $915$ & $3744$ & $28.4$  & $5.6$ & $0.20$ & $0$ & $256\times256\times129$ \\ 
  $1.08$ & $3130$ & $880$ & $922$ & $3380$ & $29.1$  & $4.5$ & $0.85$ & $1$ & $256\times256\times129$ \\ 
  $1.1$ & $3181$ & $874$ & $925$ & $3427$ & $29.3$ & $4.6$ & $0.86$ & $1$ &$256\times256\times129$ \\  
  $1.5$ & $4177$ & $887$ & $907$ & $4365$ & $28.6$ & $7.7$ & $0.91$ & $1$ &$256\times256\times129$ \\
  $2$ & $5548$ & $956$ & $910$ & $5702$ & $28.6$ & $13.1$  & $0.94$ & $1$ & $256\times512\times129$ \\
  $2.1$ & $5612$ & $1120$ & $906$ & $5792$ & $28.5$ & $12.8$  & $0.92$ & $1$ & $256\times512\times129$ \\
  $2.5$ & $6406$ & $1073$ & $901$ & $6555$ & $28.3$ & $17.6$  & $0.95$ & $1$ & $256\times512\times129$ \\
  $3$ & $4393$ & $924$ & $907$ & $4579$  & $28.5$ & $8.5$  & $0.92$ &  $2$ & $256\times512\times129$ \\
  $3$ & $4465$ & $925$ & $911$ & $4647$  & $28.7$ & $8.7$  & $0.92$ &  $2$ & $256\times768\times129$ \\
  $4$ & $5621$ & $1002$ & $894$ & $5779$ & $27.6$ & $13.9$ & $0.94$ & $2$ & $256\times512\times129$ \\
  $6$ & $4505$ & $966$ & $909$ & $4697$ & $28.5$ &  $8.9$ & $0.91$ & $4$ & $256\times1024\times129$ \\
  $8$ & $4681$ & $943$ & $900$ & $4856$ & $27.9$ & $9.7$ & $0.92$ & $5$ & $256\times1024\times129$ \\
\hline \hline
\end{tabular}
\end{center}
\caption{Simulations run at varying $L_y$ for $\Ek=10^{-5}$, $\Ra=3\times10^8$ and $L_x=1$. The value
$n$ denotes the number of positive (or, equivalently, negative) jets of the mean velocity $u_x$. The final column lists the numerical resolutions employed, with $N_i$ indicating the number of collocation points in the direction $i$.}
\label{tab:Ek1e5}
\end{table}

\subsubsection{Initial growth of the convective and large-scale flows}

\begin{figure}
\centering
       \includegraphics[clip=true,width=0.7\textwidth]{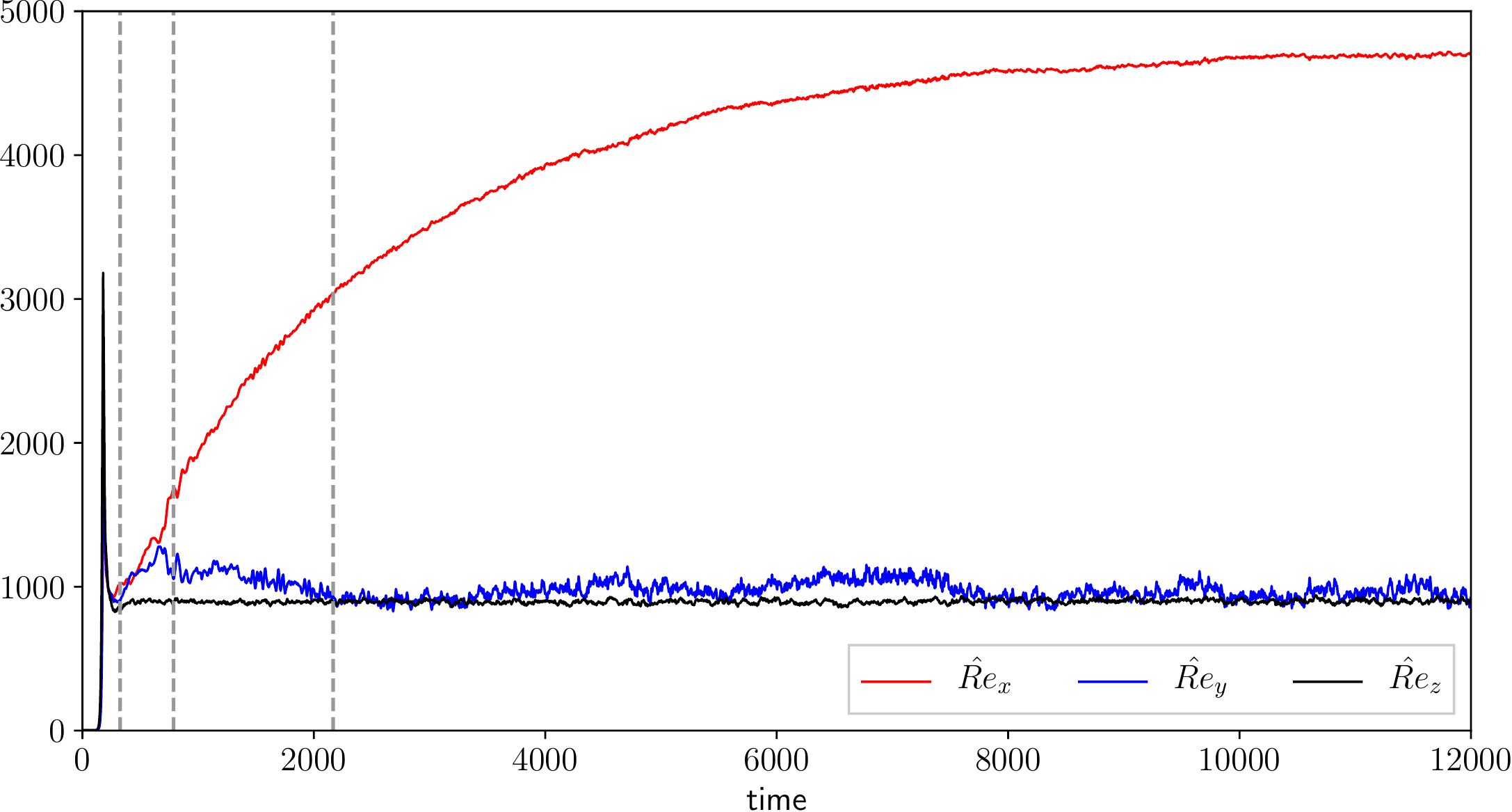}
       \caption{\label{fig:KE_Ly8}
	Time series of the r.m.s.\ values of the three velocity components for $L_y=8$. The grey dashed lines indicate the times of the snapshots in figure~\ref{fig:snap_Ly8}.}
\end{figure}

\begin{figure}
\centering
       \includegraphics[clip=true,width=\textwidth]{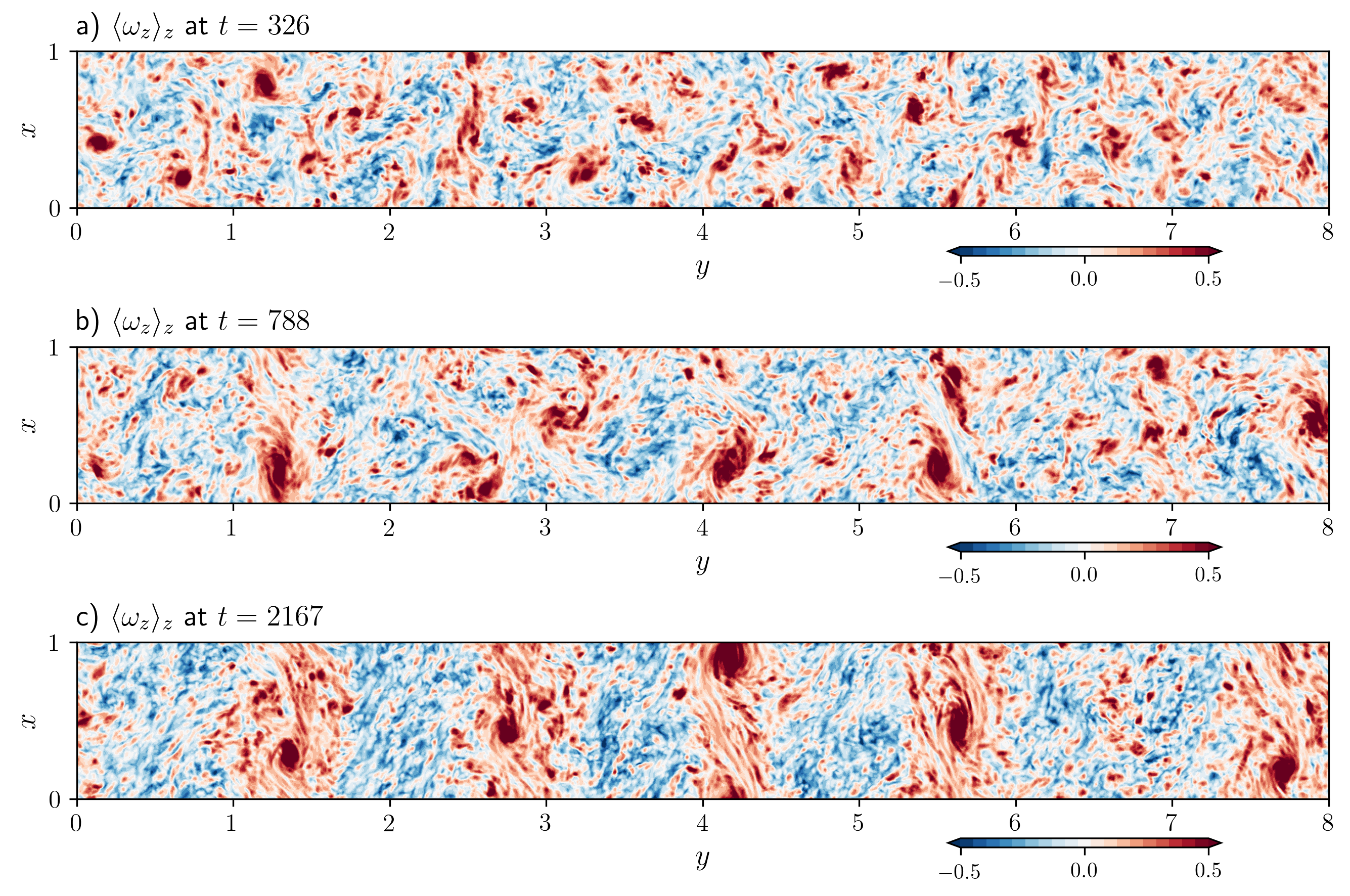}
       \caption{\label{fig:snap_Ly8}
	Snapshots of the depth-averaged axial vorticity taken at the times indicated by the grey dashed lines in figure~\ref{fig:KE_Ly8}.}
\end{figure}

Unless stated otherwise, the simulations described were started from a small perturbation to the basic state of rest. 
Figure~\ref{fig:KE_Ly8} shows the time series of $\hat{\Rey}_x$, $\hat{\Rey}_y$, and $\hat{\Rey}_z$, 
which are defined as the Reynolds number $\Rey_i$ except with only a spatial average, for $L_y=8$.
The figure shows the evolution only for $12000$ dimensionless time units, i.e.\  $0.12$ of the global viscous timescale; the full simulation was actually run for one viscous timescale.
We first describe the initial growing phase of the kinetic energy; the saturated phase is detailed in the following subsections.
Initially, the kinetic energy grows exponentially, with the velocity characterised by small horizontal scales. 
This short period of rapid growth in the kinetic energy is followed by an abrupt fall once the nonlinear terms become important. At this stage the convection remains small scale
with a horizontal scale of approximately $0.1$ for the chosen parameters. Subsequently the kinetic energy grows again, this time on a much longer timescale, as the flow moves to large horizontal scales. 

Figure~\ref{fig:snap_Ly8} shows snapshots of the depth-averaged axial vorticity 
(denoted by $\langle \omega_z\rangle_z$, where  $\langle \, \cdot \, \rangle_i$ indicates an average along the direction $i$) at the different times indicated in figure~\ref{fig:KE_Ly8}. 
This representation focuses on the formation of large-scale depth-invariant flows;
the convective flows, whose axial vorticity is mainly anti-symmetric with respect to the mid-plane \citep{Chandrasekhar61}, are not visible on these plots.
During the period of slow evolution to large-scale flows, vortices of large horizontal scale compared with the convective scale first become established across the domain 
(figure~\ref{fig:snap_Ly8}a), with movement of the vortices in both horizontal directions. The growth of these vortices corresponds to an increase in both $\hat{\Rey}_x$ and $\hat{\Rey}_y$, 
with $\hat{\Rey}_z$ saturating at a lower value; these large-scale vortices consist essentially of depth-invariant horizontal flows.
Vortex mergers occur and the LSVs grow to a scale comparable with the shortest horizontal box size (figure~\ref{fig:snap_Ly8}b). 
Subsequently, a number of jets become established, parallel to the shortest horizontal direction (figure~\ref{fig:snap_Ly8}c). 
The LSVs persist in the presence of the jets and, in the saturated state, move only with the jets. The growth of kinetic energy is eventually saturated on a timescale that is long compared with
the dynamical timescale of the convection but short compared with the viscous timescale.

\subsubsection{Flow in the saturated phase for $L_y=1$}

\begin{figure}
\centering
       \includegraphics[clip=true,width=0.8\textwidth]{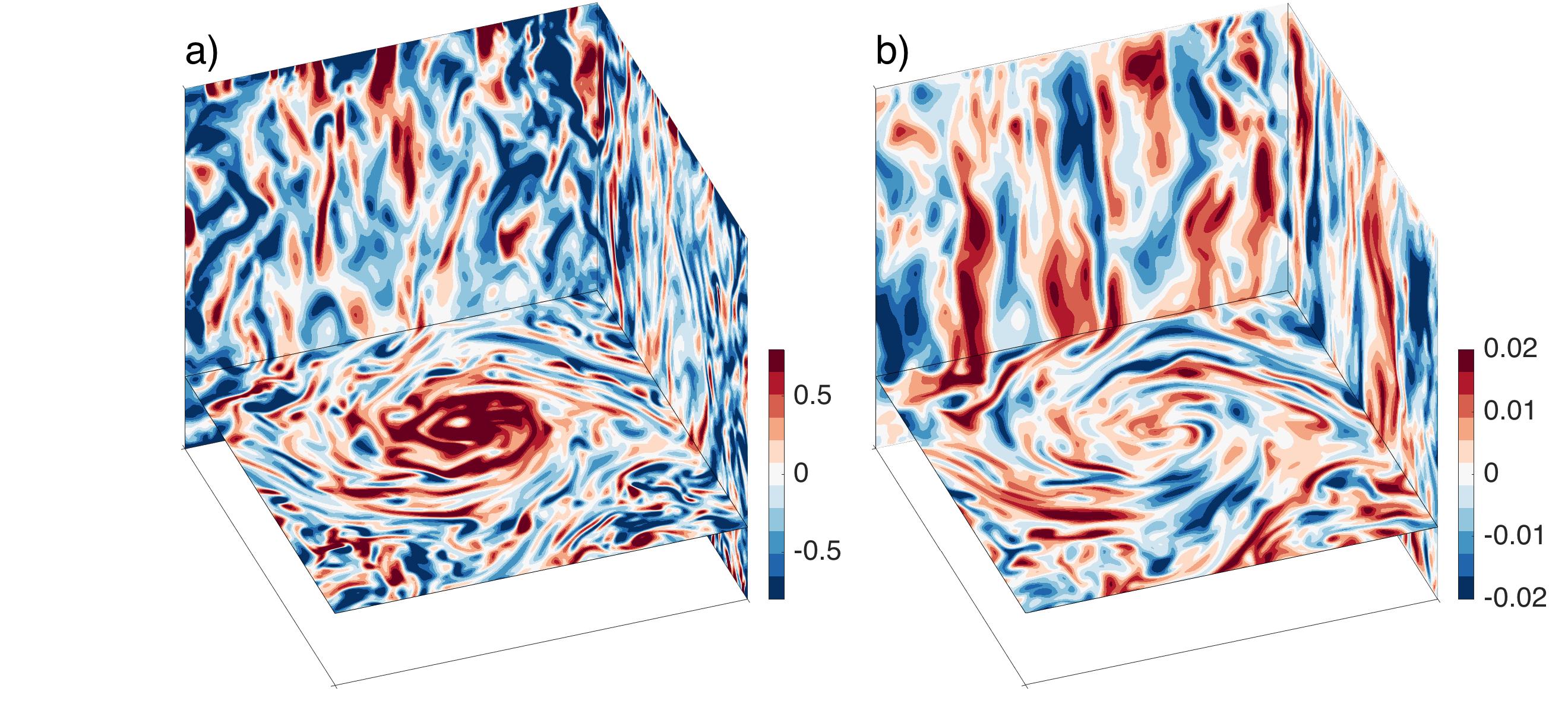}
       \caption{\label{fig:uzwz_Ly1}
	Snapshots of the horizontal and vertical cross-sections of (a) the axial vorticity and (b) the vertical velocity for $L_y=1$.}
\end{figure}

We now describe the flow in the final saturated state, first recalling results obtained for the standard case of $L_y=1$. 
Figure~\ref{fig:uzwz_Ly1} shows snapshots of horizontal and vertical cross-sections of $\omega_z$ and $u_z$.
The dominant feature in the plot of $\omega_z$ is a large-scale vortex that is predominantly $z$-invariant. 
The LSV consists of a concentrated cyclone and a more dilute anticyclone, which grow until they reach the box size. 
LSVs are a robust feature of turbulent, rotationally-constrained convection in Cartesian geometry
when $L_x=L_y$ \citep{Chan07,Kapyla11,Favier2014,Guervilly2014,Rubio14,Stellmach2014}. 
The asymmetry between cyclone and anticyclone 
is thought to be due both to the instability of the large-scale anticyclone, whose local vorticity is comparable with the planetary vorticity $2\Omega$,
and to the preference for cyclonic convective vortices \citep{Guervilly2014}. This asymmetry eventually disappears as the 
Rossby number is lowered via a decrease of the Ekman number \citep{Jul12,Stellmach2014}.

In spectral space, the LSV is a superposition of the wavenumbers 
$(k_x,k_y)=(0,1)$ and $(k_x,k_y)=(1,1)$ for the velocity component $u_x$ and of $(k_x,k_y)=(1,0)$ and $(k_x,k_y)=(1,1)$ for  $u_y$. 
The LSV therefore has a visible signature in snapshots of $\langle u_x\rangle_{xz}$ and $\langle u_y\rangle_{yz}$. 
However, the LSV drifts in random horizontal directions over time, so long time-averages
of either $\langle u_x\rangle_{xz}$ or $\langle u_y\rangle_{yz}$ produce a zero profile. 
In this sense, the LSVs do not produce any persistent unidirectional flow.

The $u_z$ plot of figure~\ref{fig:uzwz_Ly1} shows that small-scale convective flows are embedded within the large-scale cyclone.
The convective structures can be elongated in one horizontal direction, especially in the regions of intense shear surrounding the LSV. There is no 
discernable box-size vertical flow associated with the presence of the LSV, so LSVs do not directly contribute to the outward heat transport, although they might 
have a feedback effect \citep{Jul12,Guervilly2014}.

\subsubsection{Flow in the saturated phase for $L_y>1$}

\begin{figure}
\centering
       \includegraphics[clip=true,width=\textwidth]{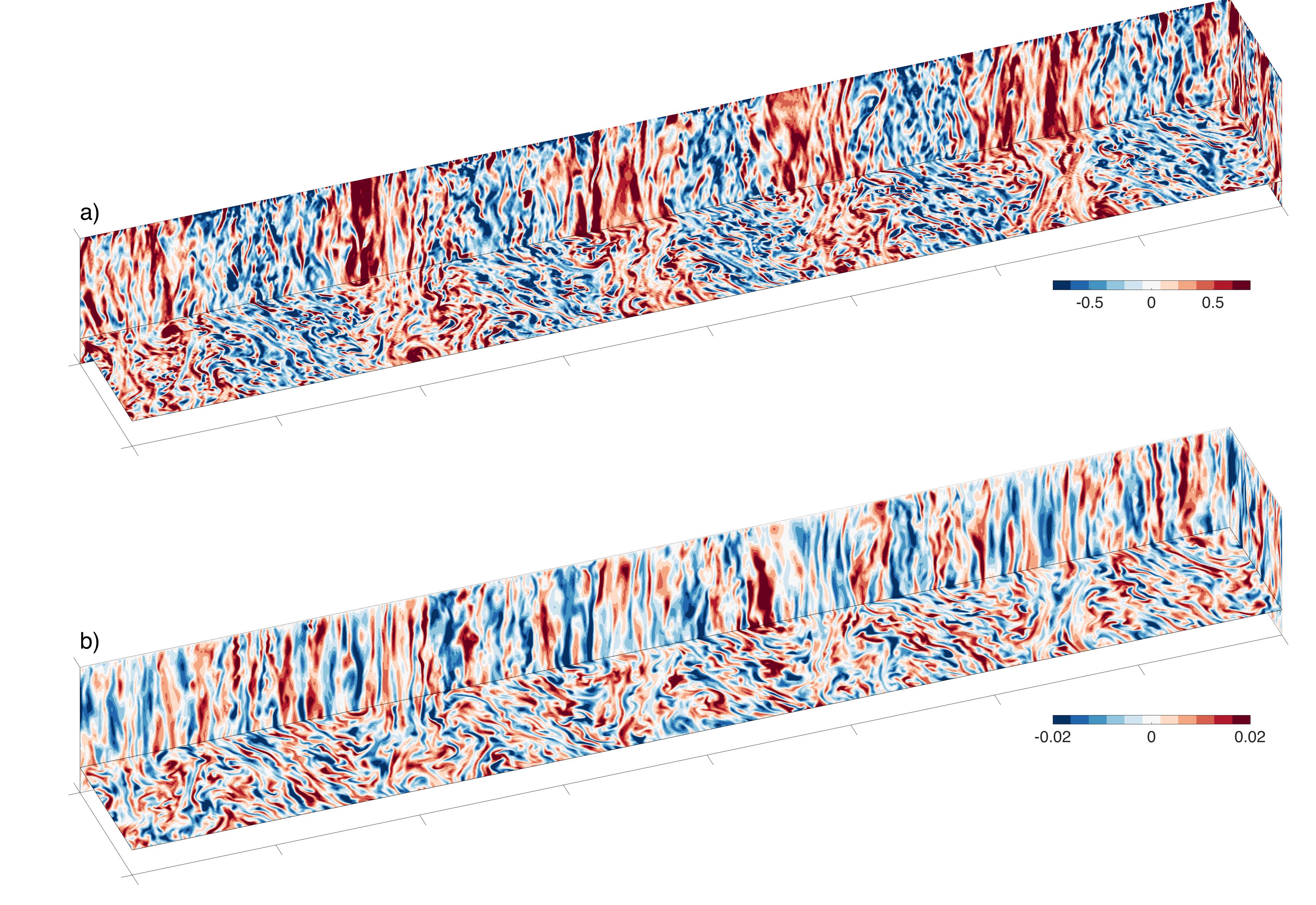}
       \caption{\label{fig:uzwz_Ly8}
	Snapshots of the horizontal and vertical cross-sections of (a) the axial vorticity and (b) the vertical velocity for $L_y=8$.}
\end{figure}

\begin{figure}
\centering
       \includegraphics[clip=true,width=\textwidth]{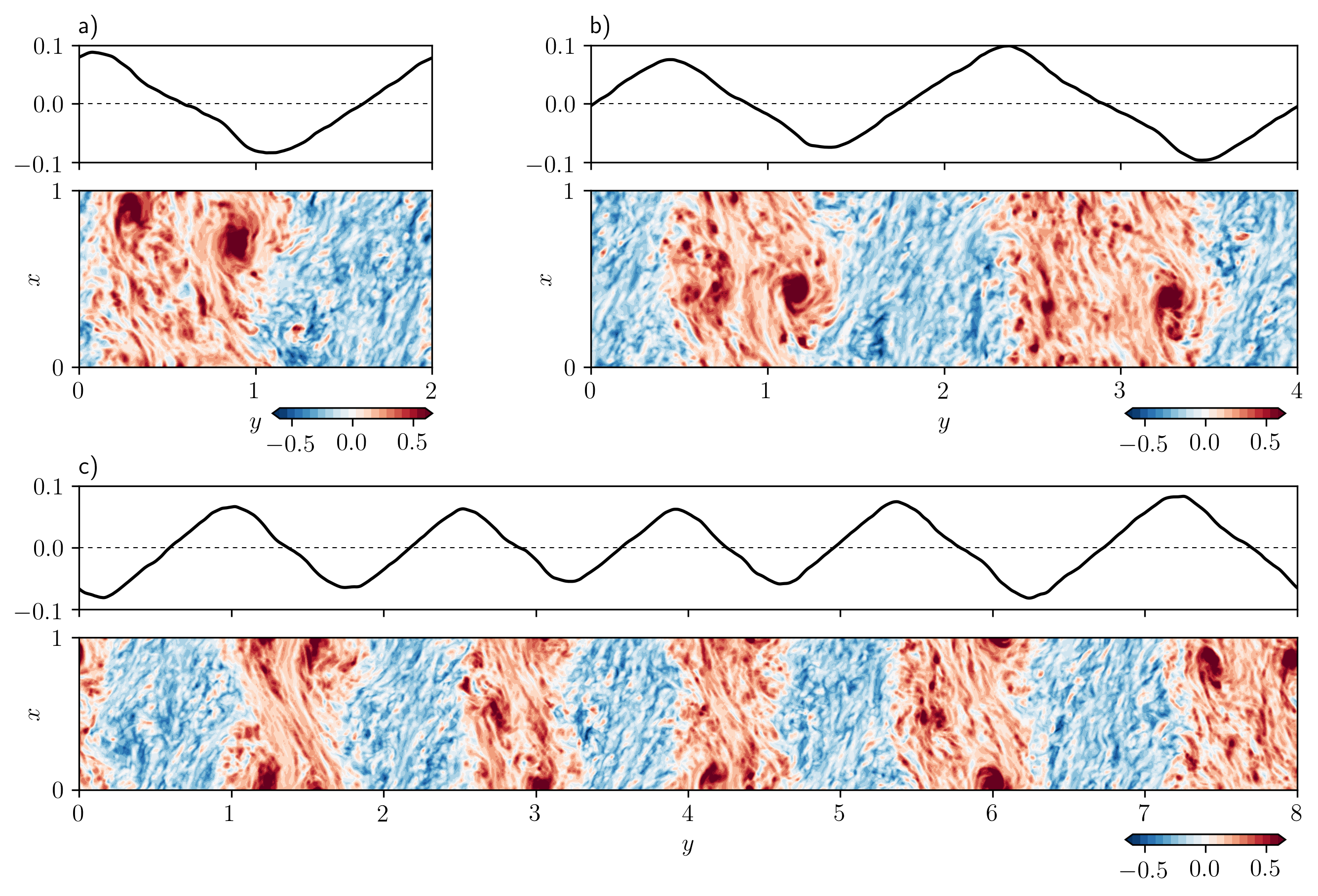}
       \caption{\label{fig:wz}
	Snapshots of the depth-averaged axial vorticity (bottom panel) and profile of the mean $u_x$ as a function of $y$ (top panel)  
	for (a) $L_y=2$, (b) $L_y=4$ and (c) $L_y=8$.}
\end{figure}

Figure~\ref{fig:uzwz_Ly8} shows snapshots of the horizontal and vertical cross-sections of $\omega_z$ and $u_z$ for $L_y=8$ in the final saturated state. 
The dominant feature in the plot of $\omega_z$ is the unidirectional flow that consists of five jets, each with a cyclonic and anti-cyclonic component, directed along $x$.
A multitude of small-scale convective vortices are embedded within the jets. The cyclonic LSVs that are prominent in the plots of  $\langle \omega_z\rangle_z$ during
the initial growing phase shown in figure~\ref{fig:snap_Ly8} are also visible in the saturated phase. 
However, as seen in figure~\ref{fig:snap_Ly8}c, they are no longer located in the middle of the bands of mean (\ie $x$-averaged) cyclonic axial vorticity, but lie towards the flanks. 
The plot of $u_z$ shows that the convective structures can be elongated along the $x$-direction in the shear regions but that there is no 
discernible box-size vertical flow associated with the presence of the jets.
Figure~\ref{fig:wz} shows snapshots of $\langle \omega_z\rangle_z$ and of the profile of $\langle u_x\rangle_{xz}$
in the saturated stage for $L_y=8$ and for the intermediate cases $L_y=2$ and $4$.
The unidirectional flow consists of one jet for $L_y=2$ and two jets for $L_y=4$. 
No persistent unidirectional flow along $y$ is produced.

The presence of a unidirectional flow might be expected in this system by analogy with 2D turbulence,
in which the upscale energy transfer feeds the largest mode available \citep[\eg][]{Bouchet2009}.
However, as noted by  \citep{Frishman2017} in a study of forced 2D turbulence,  the largest mode argument does not explain why there is
more than one stable jet for $L_y=4$ and $L_y=8$.
In all the cases studied, LSVs coexist with the jets.
The LSVs are advected by the mean flow $u_x$ and always remain at the same location in $y$. 
At most two cyclones can coexist in the same cyclonic band
and they remain in the flanks of their respective jets, with no cyclone sharing its $y$ location.   
The cyclones are never observed at the location corresponding to $\langle u_x \rangle_{xz}=0$, which is the region of maximum shear $\partial_y \langle u_x \rangle_{xz}$. Furthermore, no large-scale anticyclone is visible in the bands of anticyclonic vorticity; hence
the asymmetry between large-scale cyclones and anticyclones is present for $L_y > L_x$, as it is when $L_y = L_x$. 
The mean flows themselves however are not asymmetric, since cyclonic and anticyclonic bands have comparable width (and hence comparable amplitude).

\subsection{How much horizontal anisotropy is required to drive the jets?}
\label{sec:anisotropy}

\begin{figure}
\centering
       \includegraphics[clip=true,width=0.7\textwidth]{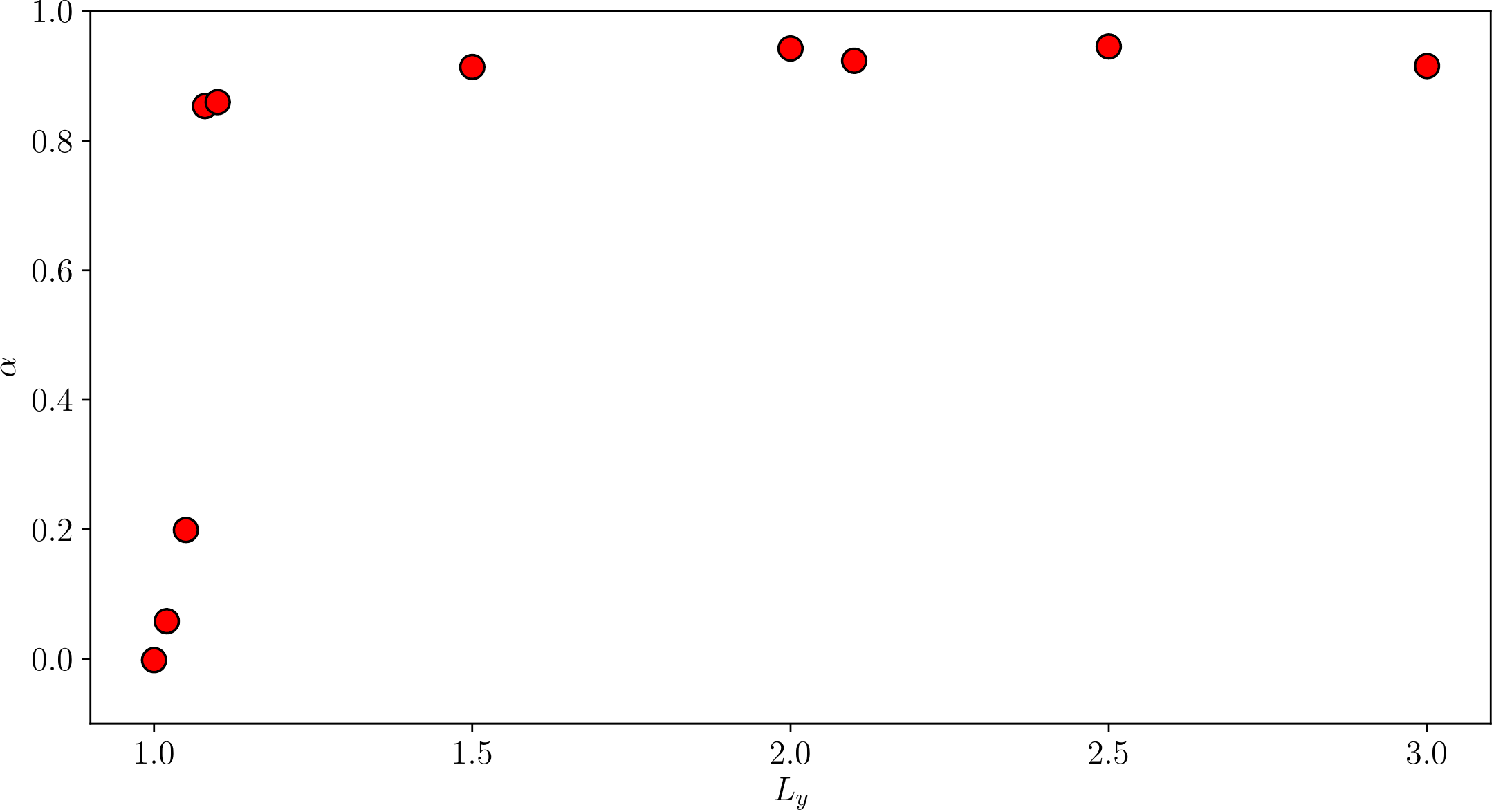}
       \caption{\label{fig:alpha}
	Flow anisotropy coefficient $\alpha$, defined by \eqref{eq:alpha}, as a function of $L_y$.}
\end{figure}

Since large-scale motions consist mostly of horizontal flows, we follow \citep{Guervilly2014} in using the parameter $\Gamma=\Rey^2/(3\Rey_z^2)$ to determine their presence; $\Gamma>1$ indicates the presence of either jets or LSVs. Such is the case for all the simulations reported in table~\ref{tab:Ek1e5}.
To distinguish between those simulations in which only LSVs form and those in which jets are present, we calculate the coefficient $\alpha$ defined by
\begin{equation}
\alpha=\frac{\langle u_x^2 \rangle -  \langle u_y^2 \rangle}{\langle u_x^2 \rangle + \langle u_y^2 \rangle },
\label{eq:alpha}
\end{equation}
which measures the anisotropy of the flow; $\alpha$ is close to unity when jets are present.
Figure~\ref{fig:alpha} shows $\alpha$ as a function of $L_y$; crucially, only a small deviation of 
$L_y$ from $L_x$ (namely $L_y=1.08$ for the parameters used here) is required to observe the spontaneous emergence of jets.

\subsection{Long term evolution and merging of the jets}
\label{sec:merging}

\begin{figure}
\centering
       \includegraphics[clip=true,width=0.9\textwidth]{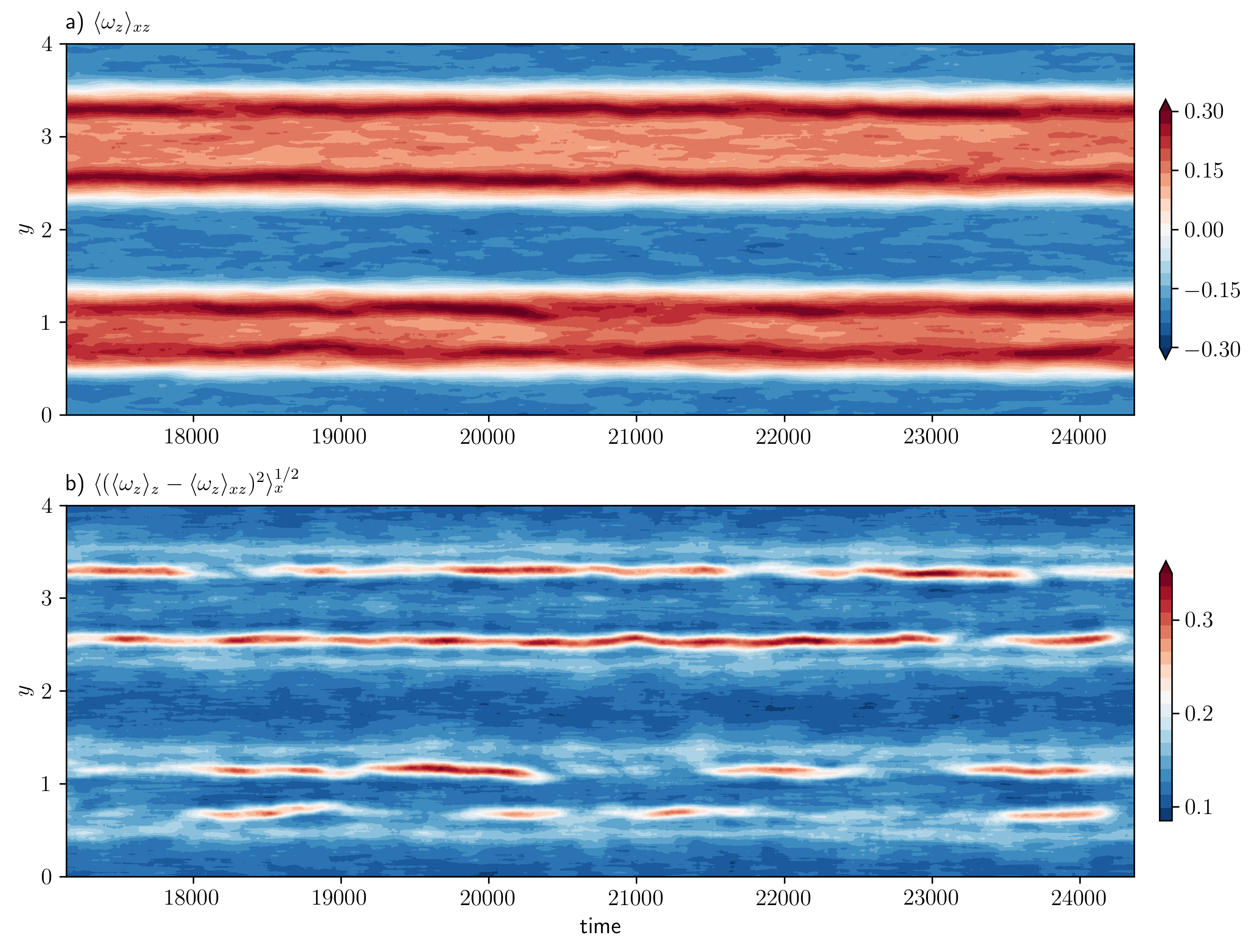}
       \caption{\label{fig:drift_Ly4}
	Space-time diagram of the depth-averaged axial vorticity for $\Ek=10^{-5}$, $\Ra=3\times10^8$, $L_x=1$ and $L_y=4$:
	(a) $x$-averaged, and (b) the r.m.s.\ of the fluctuating component.}
\end{figure}

To study the drifting or merging of the jets and LSVs over time, we plot in figure~\ref{fig:drift_Ly4} the space-time diagram of the axial vorticity averaged in $x$ and $z$,
$\langle \omega_z\rangle_{xz}$ (the mean vorticity), together with the r.m.s.\ of the fluctuating component, defined as $\langle \omega_z\rangle_{z} - \langle \omega_z\rangle_{xz}$.
The mean vorticity contains information about both the jets and the LSV, while the fluctuating part contains information about the LSV. 
Over this time series, the jets do not drift or merge; indeed, this is representative of all the cases of table~\ref{tab:Ek1e5}, irrespective of the number of jets. 
The absence of measurable drift is consistent with the fact that the horizontal average of the horizontal velocity is negligible in these simulations.
LSVs can persist in the same band of mean cyclonic vorticity for the whole duration of the time integration or they might be intermittent. 
LSVs on opposite flanks of the same cyclonic band are advected in opposite directions and do not appear to interact with each other. 
However, in some occurrences, the formation of a new LSV might destabilise the neighbouring LSV, since the disappearance of an LSV sometimes closely follows the appearance
of its immediate neighbour.
It is plausible that the persistence or intermittency of the LSVs is linked to the width of the band of mean vorticity in which they are located. Indeed, in figure~\ref{fig:drift_Ly4},
the LSVs within the narrower band located around $y=1$ are more intermittent than those within the wider band located around $y=3$. 
The comparison of the behaviour of the LSVs for the cases $L_y=2$ and $2.5$ (for which one jet is present) also supports this argument.

Table~\ref{tab:Ek1e5} gives the number of jets, denoted by $n$, where, by jets, we mean unidirectional flows that persist in time averages. For clarification, we reiterate that we are defining a jet as consisting of both its cyclonic and anticyclonic component (i.e.\ both a red and a blue region in plots such as figure~\ref{fig:snap_Ly8}). Note that simulations that produce only LSVs (\ie $L_y<1.08$) have $n=0$; cases with $L_y\geq3$ have multiple jets ($n>1$). This result is somewhat surprising since, in a system where there is no gradient of the planetary rotation (\ie no $\beta$-effect),  one might expect the upscale energy transfer to persist until the largest available scale is attained. Over the entire duration of the simulations with $\Ek=10^{-5}$ there is no evidence of a trend for any slow drift or merger of the jets. That said, for such a small value of $\Ek$, computational limitations restrict the time integration to less than one viscous timescale (which corresponds to $\Ek^{-1}$ in our time unit).
In all the simulations reported so far, we have described the large-scale structure that gradually emerges from an initial small-scale perturbation and that persists over the entire  
time integration. 
The system spontaneously adopts a number of jets such that $L_y/(2n)\approx L_x$ when possible (\ie when $L_y$ is large enough). This observation can be explained by
the snapshots of the initial growing stage in figure~\ref{fig:snap_Ly8}: the LSVs form first and grow to the size $L_x$, before extending into jets that retain
the initial size of the LSV in the $y$-direction.
It is however plausible that on longer timescales the jets could readjust their size by merging. 

\begin{figure}
\centering
       \includegraphics[clip=true,width=0.7\textwidth]{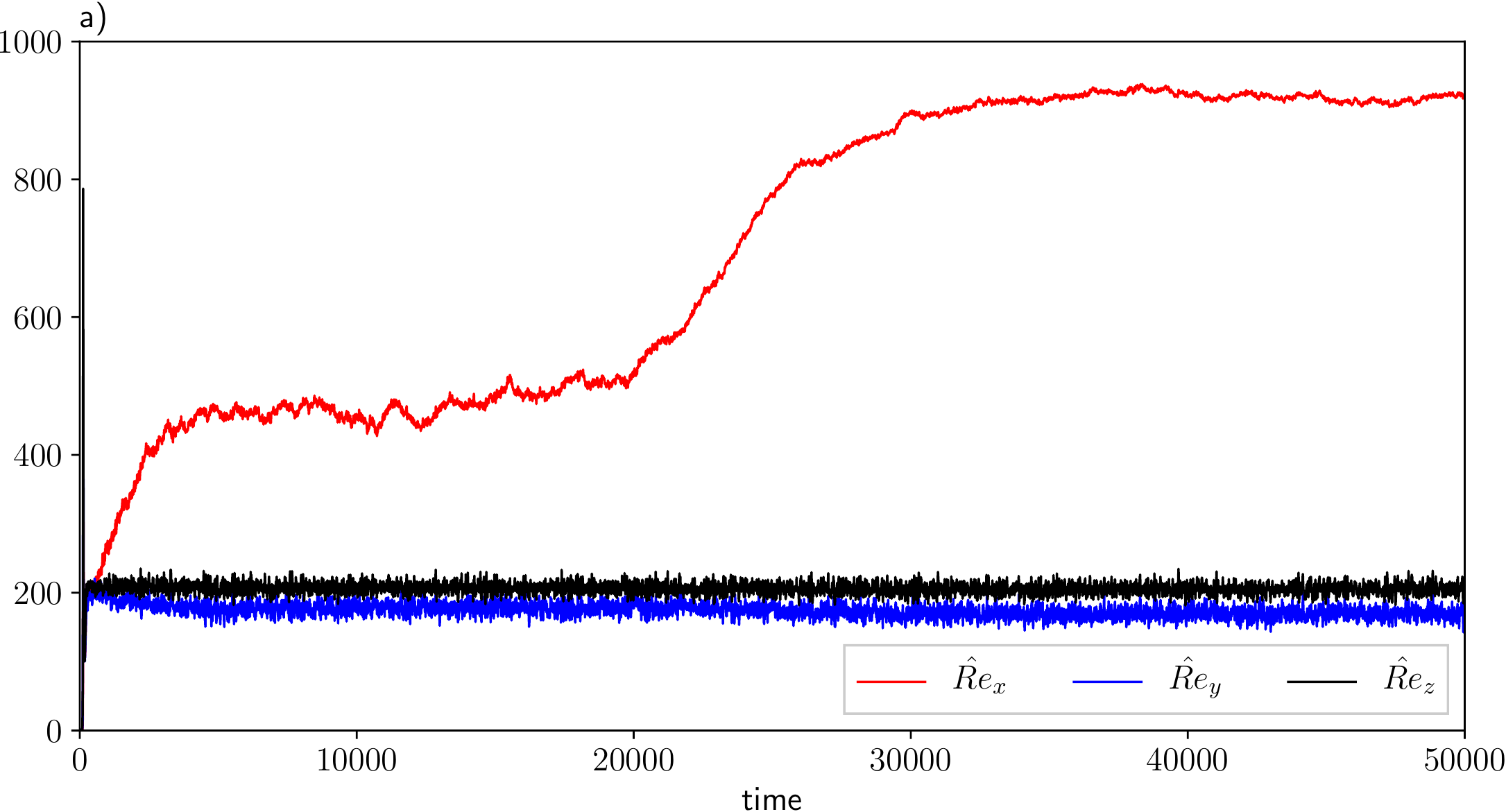}
       \includegraphics[clip=true,width=0.9\textwidth]{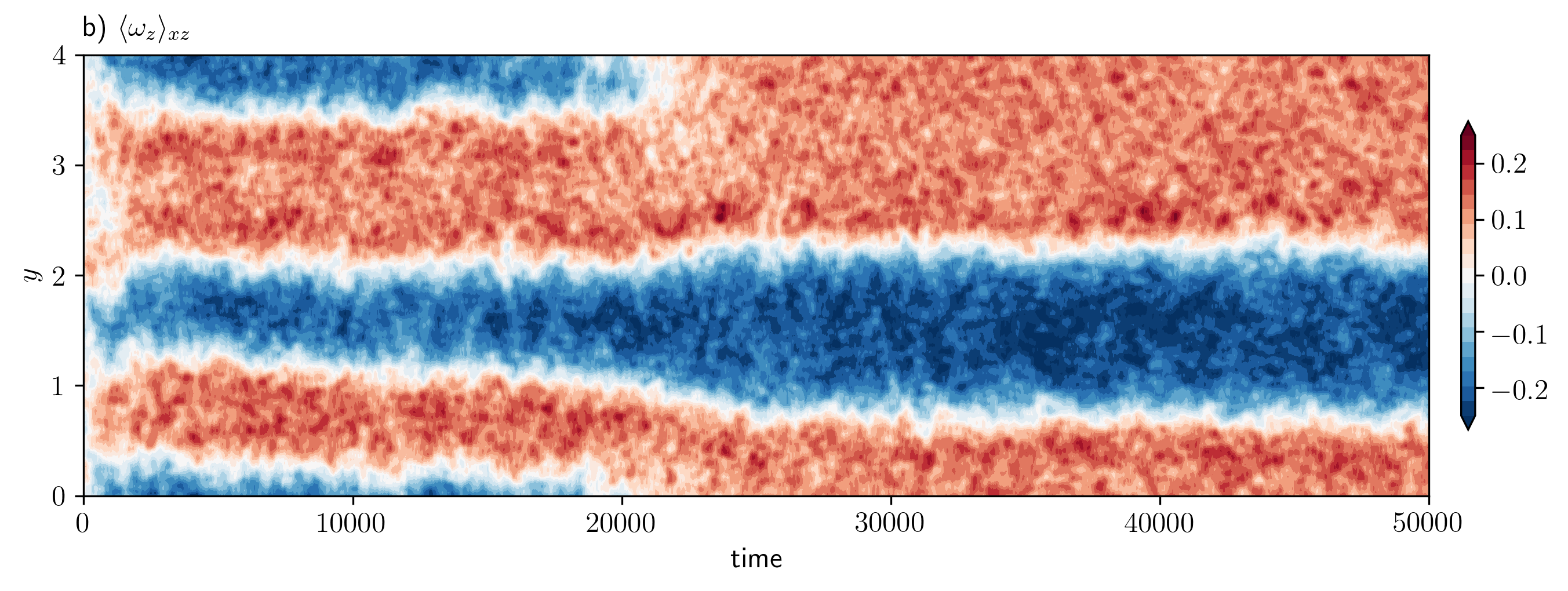}
       \caption{\label{fig:merger_Ly4}
        a) Time series of the r.m.s.\ values of the three velocity components and 
	b) space-time diagram of the depth-averaged axial vorticity for $\Ek=10^{-4}$, $\Ra/\Ra_c=4.2$ and $(L_x,L_y)=(1,4)$.}
\end{figure}

\begin{figure}
\centering
       \includegraphics[clip=true,width=0.7\textwidth]{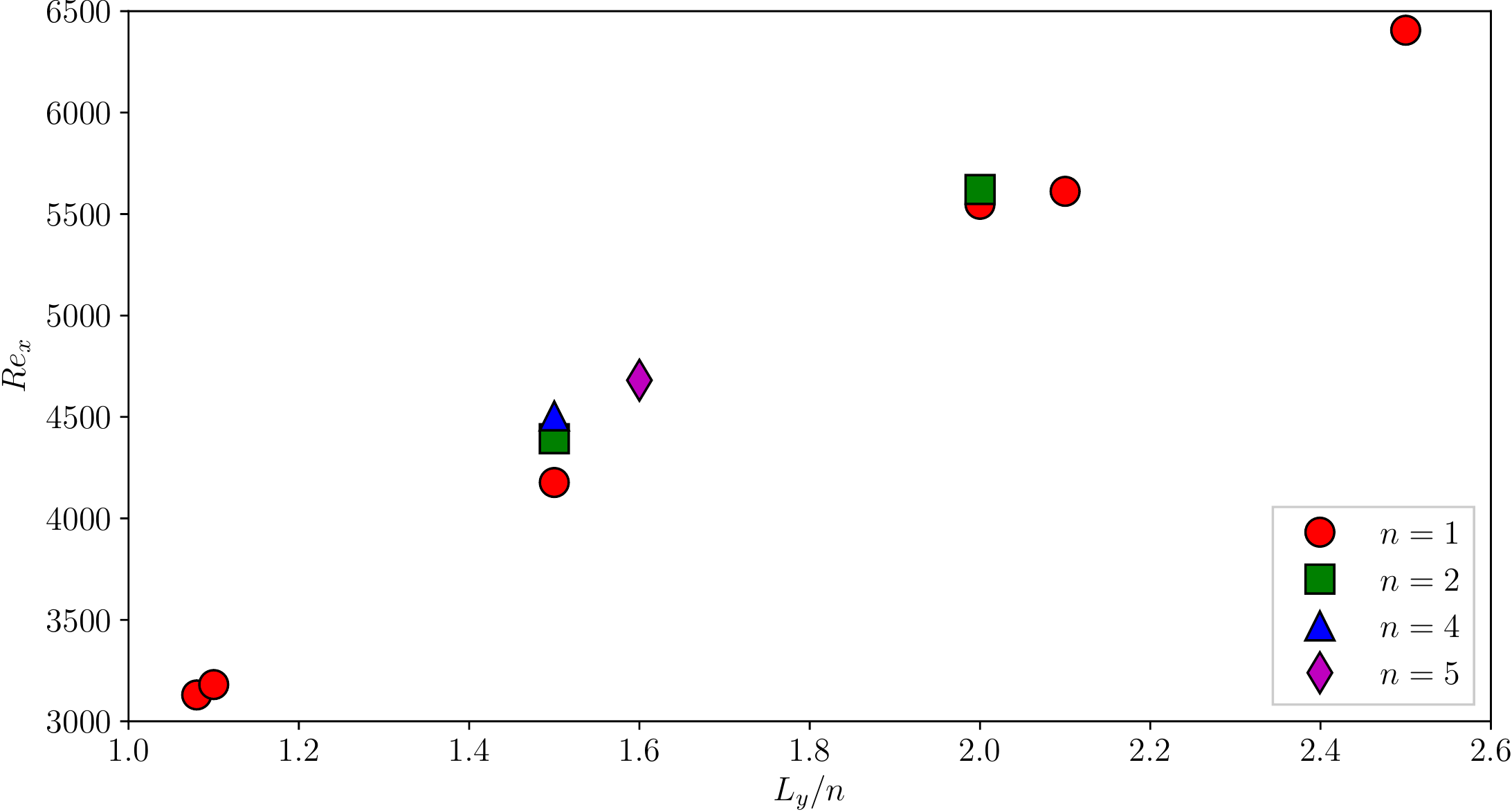}
       \caption{\label{fig:RexLy}
	$\Rey_x$ as a function of $L_y/n$ (for cases with $n>0$) for $\Ek=10^{-5}$, $\Ra=3\times10^8$ and $L_x=1$.}
\end{figure}

In order to investigate very long term aspects of the dynamics, it is necessary to consider an increased value of the Ekman number. To this end, we perform a simulation
with $\Ek=10^{-4}$, $\Ra=8\times10^{6}$ ($\Ra/\Ra_c=4.2$) and $(L_x,L_y)=(1,4)$ for a duration of five viscous timescales.
The time series of $\hat{\Rey}_x$, $\hat{\Rey}_y$ and $\hat{\Rey}_z$, together with the space-time diagram of $\langle \omega_z\rangle_{xz}$, are 
shown in figure~\ref{fig:merger_Ly4}. The initial phases of growth and saturation, before one viscous timescale has elapsed, are similar to the case already described for 
$\Ek=10^{-5}$, with the formation of four jets. 
Note that no persistent LSV is observed in the jet flanks here because $L_x$ is not large enough to permit it at this value of $\Ek$.
Over approximately two viscous timescales, one of the mean cyclonic bands slowly drifts, preceding a relatively rapid expansion and merger, with only one jet remaining; this jet merger is
accompanied by an increase by a factor two of $\hat{\Rey}_x$.
On long timescales, the jets therefore evolve by increasing their lengthscale as a result of the merger of bands of mean cyclonic vorticity, leading to an increase in the amplitude of $u_x$.
This relation between the amplitude of $u_x$ and the jet lengthscale is consistent with the results presented in table~\ref{tab:Ek1e5} (for which $\Ek=10^{-5}$).
Figure~\ref{fig:RexLy} shows $\Rey_x$ as a function of the jet lengthscale $L_y/n$ for the cases with $n>0$ of table~\ref{tab:Ek1e5}, corroborating the linear
dependence of $\Rey_x$ on $L_y/n$.

\subsection{Bistability of the large-scale flow}
\label{sec:bistability}

In their study of forced 2D turbulence, \citet{Bouchet2009} show that when the aspect ratio of the computational domain slightly exceeds unity, the flow switches randomly between a dipole vortex and a unidirectional flow. For the ratio $L_y/L_x=1.02$, their system spends more time 
in the dipole vortex configuration, in contrast with the case of $L_y/L_x=1.04$, for which the unidirectional flow configuration is preferred. These topological changes are found to be slow processes that occur over a viscous timescale, and which therefore can be observed only through long time integrations.  
In all our simulations reported in table~\ref{tab:Ek1e5} (all of which have $Ek=10^{-5}$), which are started from an initial small-scale perturbation, no random topological changes were observed, even for small deviations of $L_y/L_x$ from unity. However, the results of \citep{Bouchet2009} show that 
the solution might be dependent on initial conditions and may vary on viscous timescales.

\begin{figure}
\centering
       \includegraphics[clip=true,width=0.7\textwidth]{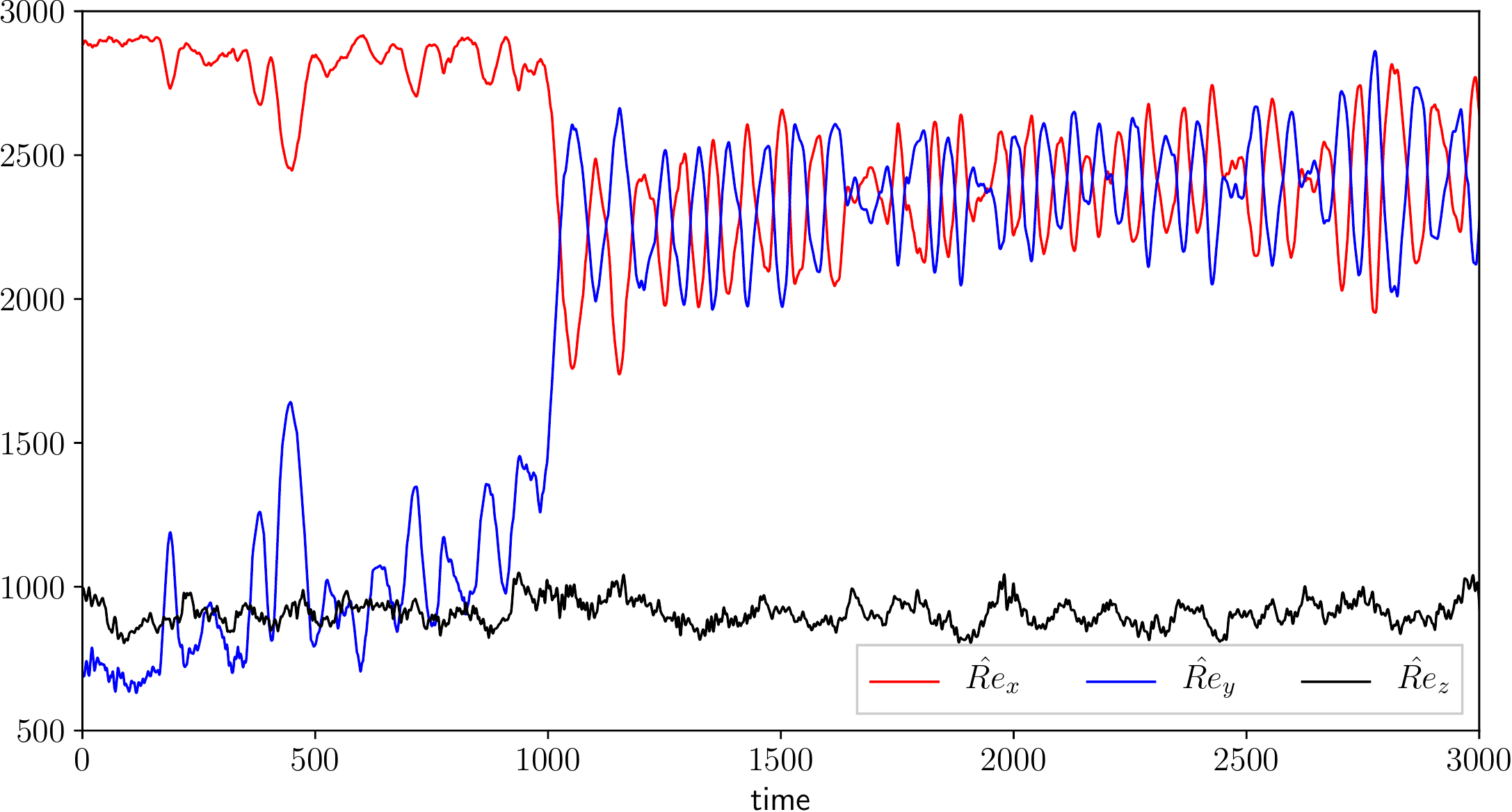}
       \caption{\label{fig:stab1a}
       Time series of the r.m.s.\ values of the three velocity components for the simulation with $(L_x,L_y)=(1,1)$ whose initial condition is constructed from $(L_x,L_y)=(0.5,1)$
       ($\Ek=10^{-5}$, $\Ra=3\times10^8$).}
\end{figure}

To study whether the two configurations --- LSV-only or jets --- can both occur for a given aspect ratio $L_y/L_x$,
we should ideally perform simulations for several viscous timescales. 
Since computational constraints exclude this possibility for small Ekman numbers, we use two different approaches.
First, we force the system to adopt a given large-scale configuration by using an initial condition of finite amplitude.
To determine whether the jets can be stable when $L_x=L_y$, we use the data from a snapshot of the case $(L_x,L_y)=(0.5,1)$ (which is thus in the jet configuration), which we replicate in the $x$-direction to obtain a box of dimension $(L_x,L_y)=(1,1)$. This fabricated data, with the addition of a small amount of noise, is then used as an initial condition for the domain $(L_x,L_y)=(1,1)$. 
Figure~\ref{fig:stab1a} shows the time series of $\hat{\Rey}_x$, $\hat{\Rey}_y$ and $\hat{\Rey}_z$. 
The jet configuration is recognisable initially by $\hat{\Rey}_x>\hat{\Rey}_y\approx\hat{\Rey}_z$.  
After about 1000 time units, or equivalently about 25 turnover timescales based on the horizontal velocity and the box size, 
the jet configuration is lost in favour of the LSV configuration (where $\hat{\Rey}_x\approx\hat{\Rey}_y$). 
Furthermore, we perform simulations for which a snapshot from a case with $(L_x,L_y)=(1,1.08)$ (a jet configuration) is used as an initial condition for either $(L_x,L_y)=(1,1)$ or 
$(L_x,L_y)=(1.08,1.08)$. In both cases, the large-scale flow rapidly evolves from jets to an LSV, similarly to the case shown in figure~\ref{fig:stab1a}.
Based on these numerical simulations, we conclude that the jets become rapidly unstable when $L_x=L_y$.

\begin{figure}
\centering
       \includegraphics[clip=true,width=0.7\textwidth]{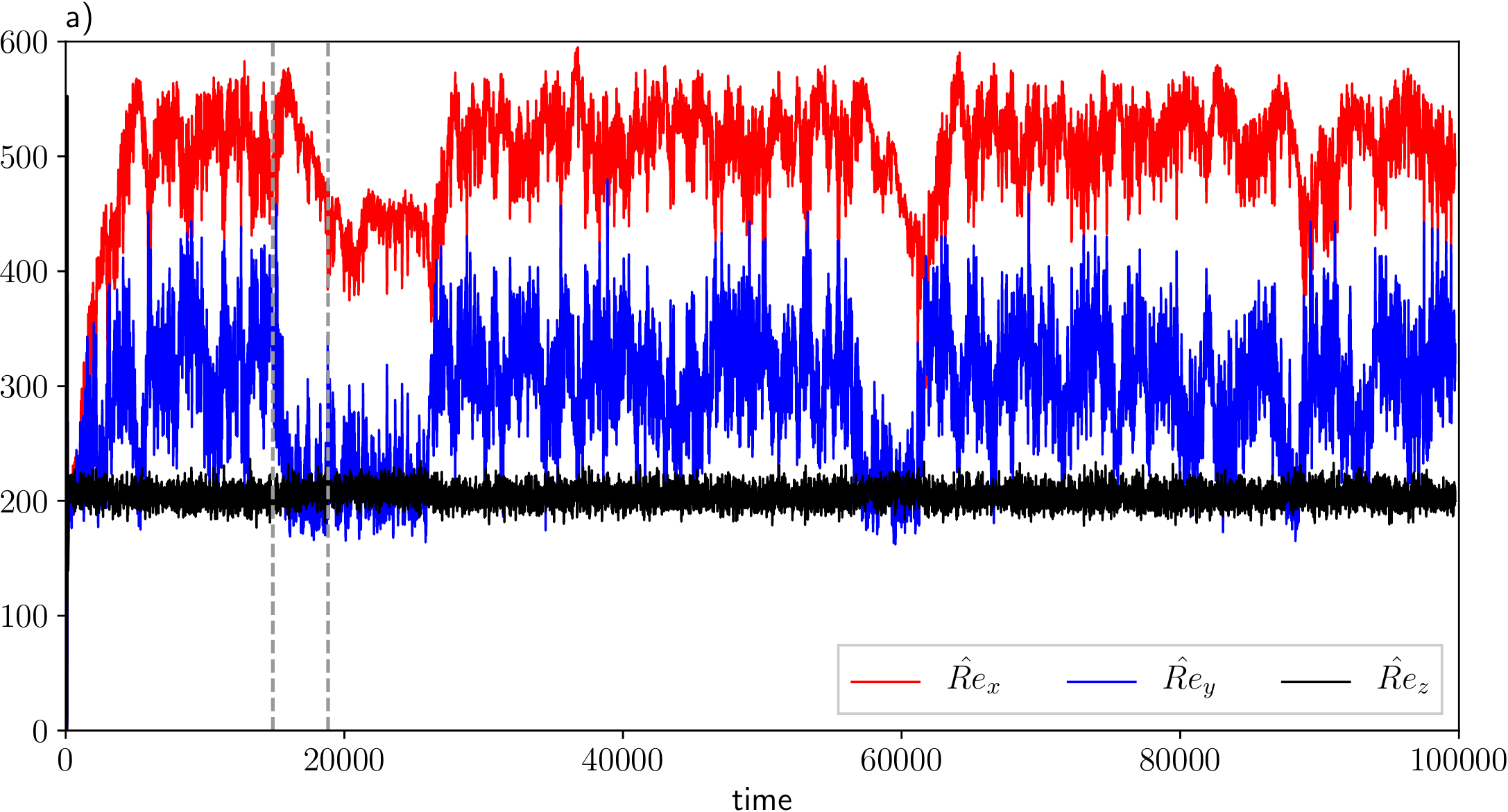}
       \includegraphics[clip=true,width=0.7\textwidth]{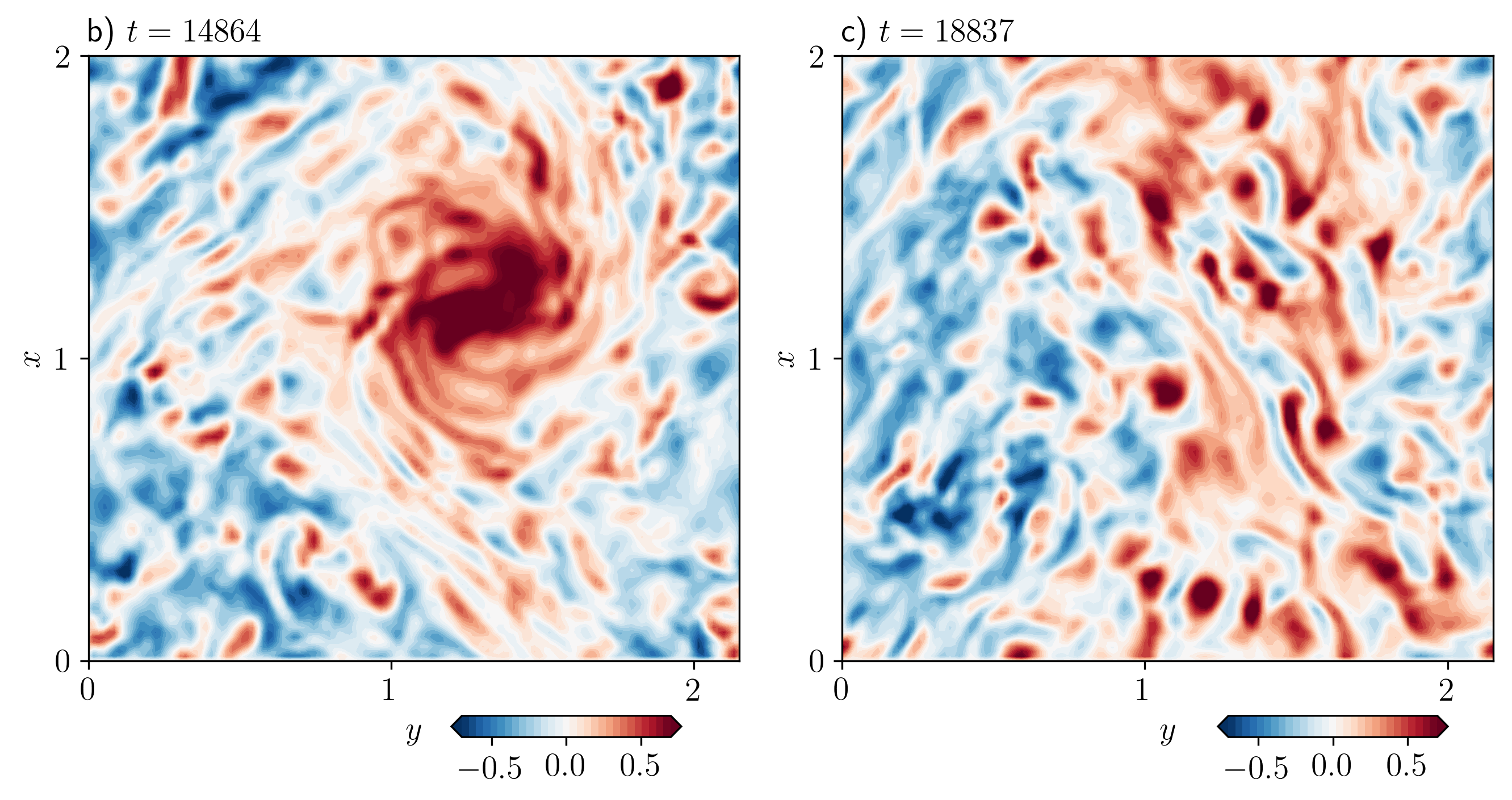}
       \caption{\label{fig:stab2}
	(a) Time series of the r.m.s.\ values of the three velocity components for the simulation with $\Ek=10^{-4}$, $\Ra/\Ra_c=4.2$ and $(L_x,L_y)=(2,2.15)$.
	(b)--(c) Snapshots of the $z$-averaged axial vorticity at the times indicated by the dashed vertical lines in (a).}
\end{figure}

\begin{figure}
\centering
       \includegraphics[clip=true,width=0.7\textwidth]{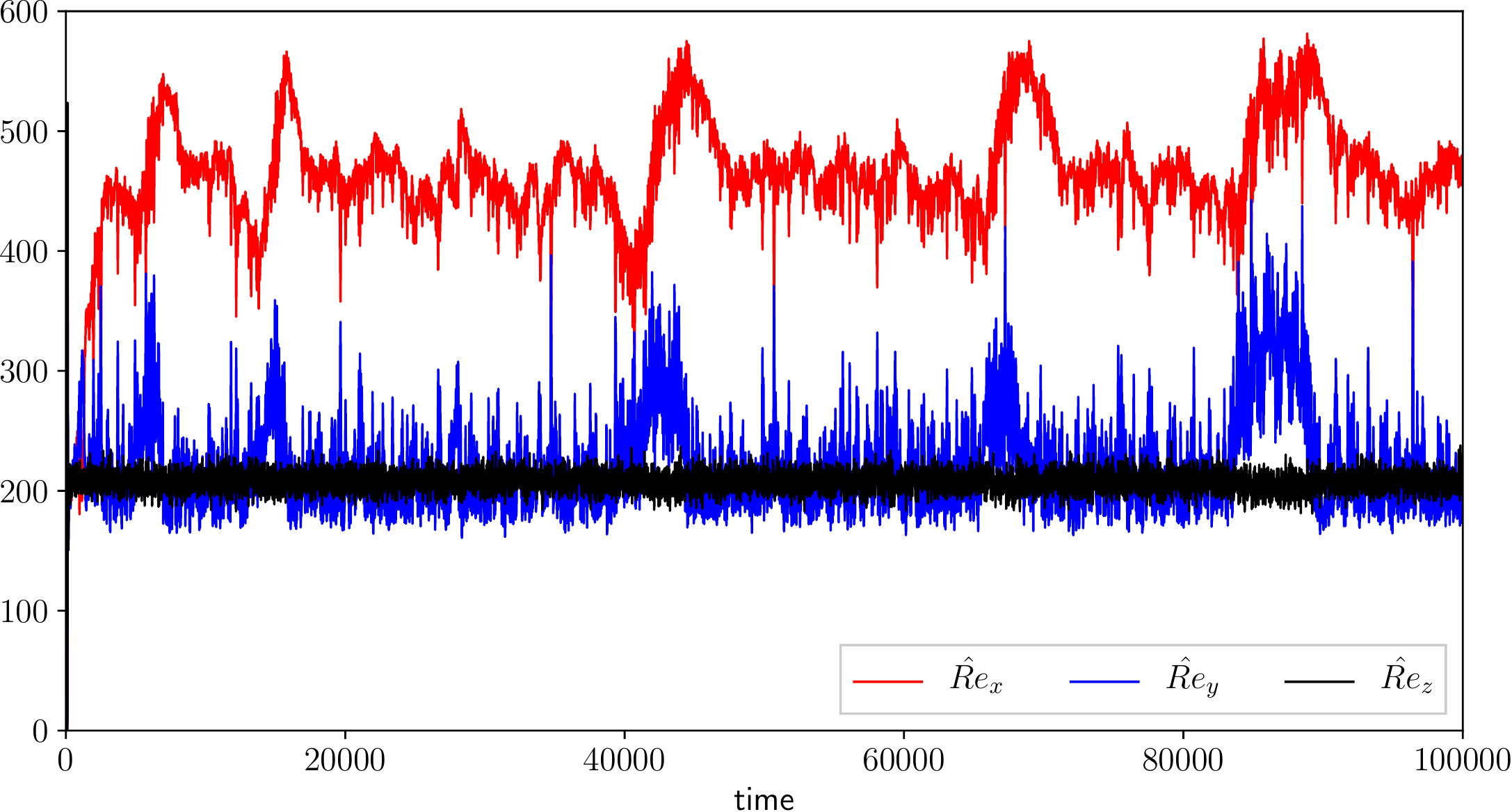}
       \caption{\label{fig:stab3}
	Time series of the r.m.s.\ values of the three velocity components for the simulation with $\Ek=10^{-4}$, $\Ra/\Ra_c=4.2$ and $(L_x,L_y)=(2,2.2)$.}
\end{figure}

The second approach consists, as in \S\,\ref{sec:merging}, in increasing the Ekman number in order to simulate many viscous timescales. We perform a simulation
with $\Ek=10^{-4}$, $\Ra/\Ra_c=4.2$ and $(L_x,L_y)=(2,2.15)$ for a duration of $10$ viscous timescales.
The time series of $\hat{\Rey}_x$, $\hat{\Rey}_y$ and $\hat{\Rey}_z$ are shown in figure~\ref{fig:stab2}a. 
An LSV forms initially, followed by spontaneous topological changes that occur after the elapse of more than one viscous timescale, which corresponds to $10^4$ time
units in figure~\ref{fig:stab2}a.
The LSV configuration is characterised by $\hat{\Rey}_y>\hat{\Rey}_z$  
and is illustrated in figure~\ref{fig:stab2}b, which shows a snapshot of $\langle \omega_z\rangle_{z}$.
The jet configuration is characterised by $\hat{\Rey}_y\approx\hat{\Rey}_z$ and is illustrated by figure~\ref{fig:stab2}c.
For this particular aspect ratio ($L_y/L_x=1.075$), the system spends most of its time in the LSV configuration, i.e.\ the probability of obtaining the LSV configuration is higher.  As $L_y/L_x$ decreases towards unity, the probability of a jet configuration decreases. Conversely, as $L_y/L_x$ increases, the probability of achieving the LSV configuration decreases.
This is illustrated in figure~\ref{fig:stab3}, which shows the times series of $\hat{\Rey}_x$, $\hat{\Rey}_y$ and $\hat{\Rey}_z$ for $(L_x,L_y)=(2,2.2)$.  
We find that for $L_y/L_x=1.2$, the system remains in the jet configuration during the whole time integration of $10$ viscous timescales.
For the parameters studied in this paragraph, there is a narrow window, given by $L_y/L_x\in[1.06,1.15]$, in which both the jet state and the LSV-only state can exist. 

\subsection{Domain of existence}
\label{sec:domain}

When $L_y=L_x$, LSVs form when the convective vortices are constrained by rotation to be anisotropic (\ie narrow in the horizontal directions and tall in the vertical)
and when they are sufficiently energetic to merge \citep{Guervilly2014}. At least for the Ekman numbers considered in \citet{Guervilly2014} ($\Ek\geq 5\times10^{-6}$), these conditions are quantified by (i) small local Rossby numbers, $\Ro_z^l=\langle u_z \rangle / (2\Omega l) \lesssim 0.1$, where $l$ is the typical horizontal lengthscale of the convection, and (ii) $\Ra/\Ra_c\gtrsim3$. To meet both criteria, the Ekman number must be suitably small.
An additional condition for the formation of LSVs is that the computational domain must be wide enough to accommodate a sufficient number of convective cells in the horizontal direction (typically more 
than four convective cells along $L_x$ or $L_y$).

\begin{table}
\scriptsize
\begin{center}
\begin{tabular}{c c c c c c c c c c c c c}
\hline \hline
$\Ek$ & $\Ra$ & $\Ra/\Ra_c$ & $L_x$ & $L_y$ & $\Rey_x$ & $\Rey_y$ &  $\Rey_z$ & $\Rey$ & $\Ro_z^l$ & $\Gamma$ & $\alpha$ & $N_x\times N_y\times N_z$
\\ \hline 
  $3.16\times10^{-4}$ & $1.7\times10^6$ & $4.1$ & $2$ & $2$ & $109$ & $110$ & $119$ & $195$ & $0.12$ & $0.9$ & $0.00$ & $128\times128\times97$ \\
  $3.16\times10^{-4}$ & $1.7\times10^6$ & $4.1$ & $1$ & $4$ & $119$ & $105$ & $119$ & $198$ & $0.12$ & $0.9$ & $0.12$ & $64\times256\times97$ \\
  $3.16\times10^{-4}$ & $1.7\times10^6$ & $4.1$ & $2$ & $4$ & $111$ & $109$ & $119$ & $196$ & $0.12$ & $0.9$ & $0.02$ & $128\times256\times97$ \\
  \hline
  $10^{-4}$ & $8\times10^6$ & $4.2$ & $2$ & $2$ & $427$ & $426$ & $202$ & $636$ & $0.091$ & $3.3$ & $0.00$ & $128\times128\times97$ \\
  $10^{-4}$ & $8\times10^6$ & $4.2$ & $1$ & $2$ & $445$ & $177$ & $208$ & $523$ & $0.092$& $2.1$ & $0.73$ & $64\times128\times97$ \\
  \hline
   $10^{-5}$ & $5\times10^7$ & $1.2$ & $1$ & $2$  & $44$ & $43$ & $45$ & $76$ & $0.0045$ & $0.95$ & $0.02$ & $128\times256\times129$ \\
   $10^{-5}$ & $1\times10^8$ & $2.5$ & $1$ & $2$   & $389$ & $266$ & $222$ & $521$ & $0.022$& $1.84$ & $0.36$ & $128\times256\times129$\\
   $10^{-5}$ & $1.5\times10^8$ & $3.7$ & $1$ & $2$   & $2139$ & $438$ & $444$ & $2228$  & $0.041$ & $8.40$ & $0.92$ & $128\times256\times129$ \\ 
   $10^{-5}$ & $3\times10^8$ & $7.4$ & $1$ & $2$ & $5548$ & $956$ & $910$ & $5702$ & $0.084$ & $13.1$  & $0.94$ & $256\times512\times129$ \\
   $10^{-5}$ & $4\times10^8$ & $9.9$ & $1$ & $2$ & $6824$ & $1095$ & $1169$ & $7010$  & $0.107$ & $12.0$ & $0.95$ & $256\times512\times129$ \\ 
   $10^{-5}$ & $6\times10^8$ & $14.8$ & $1$ & $2$ & $7708$ & $1365$ & $1551$ & $7980$  & $0.139$ & $8.82$ & $0.94$ & $256\times512\times257$  \\ 
   $10^{-5}$ & $7\times10^8$ & $17.3$ & $1$ & $2$ & $7811$ & $1573$ & $1713$ & $8150$  & $0.153$ & $7.54$ & $0.92$ & $256\times512\times257$ \\ 
\hline \hline
\end{tabular}
\end{center}
\caption{Simulations performed in order to determine the domain of existence of the large-scale flows.}
\label{tab:existence}
\end{table}

To determine whether similar constraints apply to the formation of the jets, we first performed simulations at larger Ekman numbers than considered above. 
Table~\ref{tab:existence} lists a number of outputs from simulations run at $\Ek=3.14\times10^{-4}$ and $\Ek = 10^{-4}$ for $\Ra/\Ra_c\approx4$ and for various box configurations.
The ratio $\Gamma$ is again used to indicate the formation of dominant horizontal flows (related to the formation of large-scale flows) when greater than unity,
and the coefficient $\alpha$ is used to indicate the formation of unidirectional flows when close to unity.
For $\Ek=3.14\times10^{-4}$ and all the box configurations,
the local Rossby number is $\Ro_z^l=0.12$ and $\Gamma\approx 1$; neither an LSV nor jets form because conditions (i) and (ii) cannot be met simultaneously for this Ekman number.
Note that, for this value of $\Ek$, only about three convective cells fit along a unit horizontal length; we thus have to use wide computational domains. 
When $\Ek$ is decreased to $10^{-4}$, the local Rossby number decreases to $0.09$ for this value of $\Ra/\Ra_c$. 
For all box configurations, $\Gamma>1$, i.e.\ large-scale horizontal flows are predominant.
For $L_x=L_y=2$, $\alpha = 0$, i.e.\ the large-scale flow is an LSV, as expected. For $(L_x,L_y)=(1,2)$, $\alpha>0$, reflecting the formation of jets. 
These observations are all confirmed by visual inspection of the axial vorticity (not shown here for brevity). 
These simulations thus show that the constraint $\Ek \lesssim 10^{-4}$ is required for the formation of jets, a similar criterion to that for the formation of LSVs.

\begin{figure}
\centering
       \includegraphics[clip=true,width=0.49\textwidth]{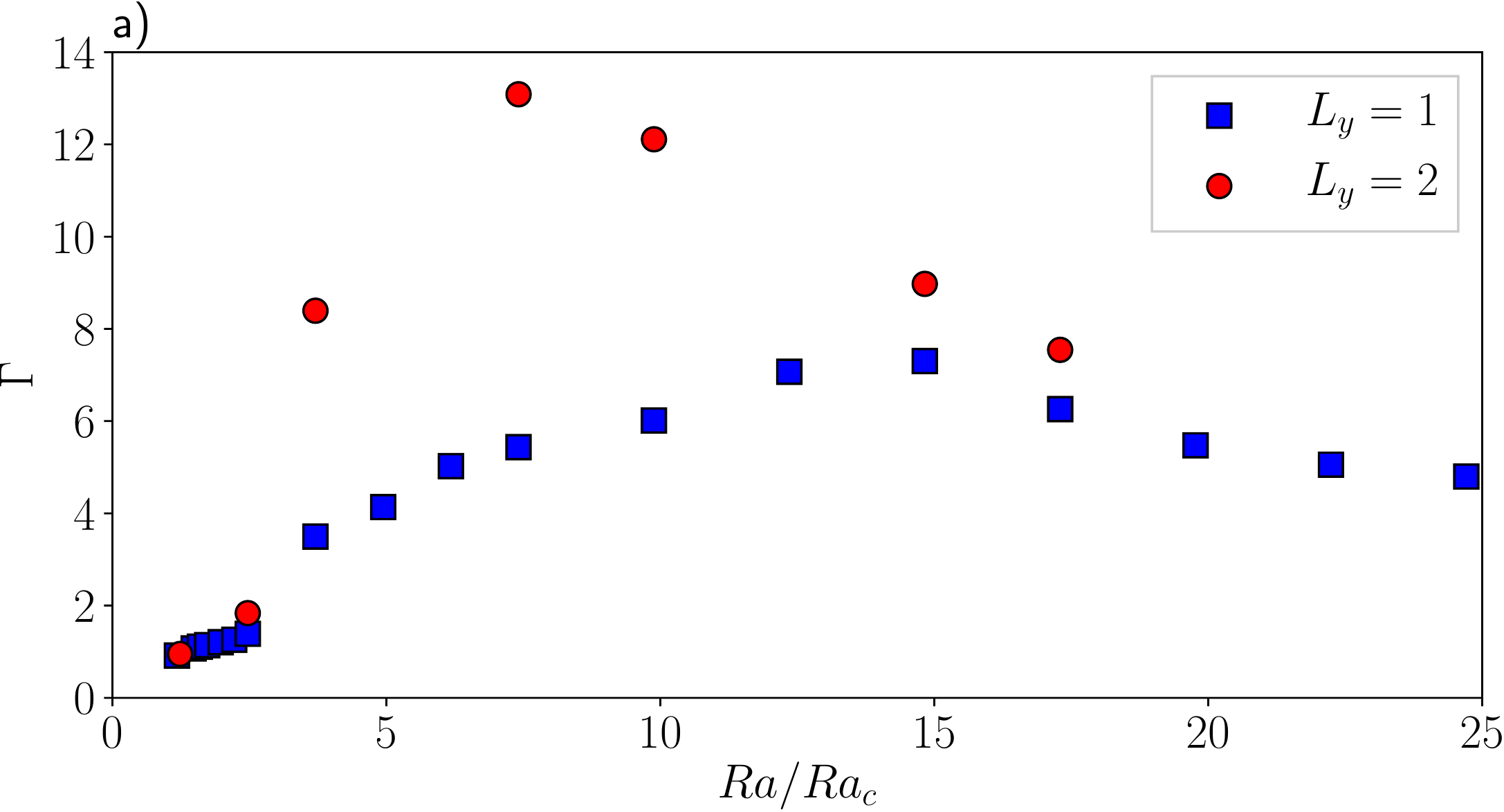}
       \includegraphics[clip=true,width=0.49\textwidth]{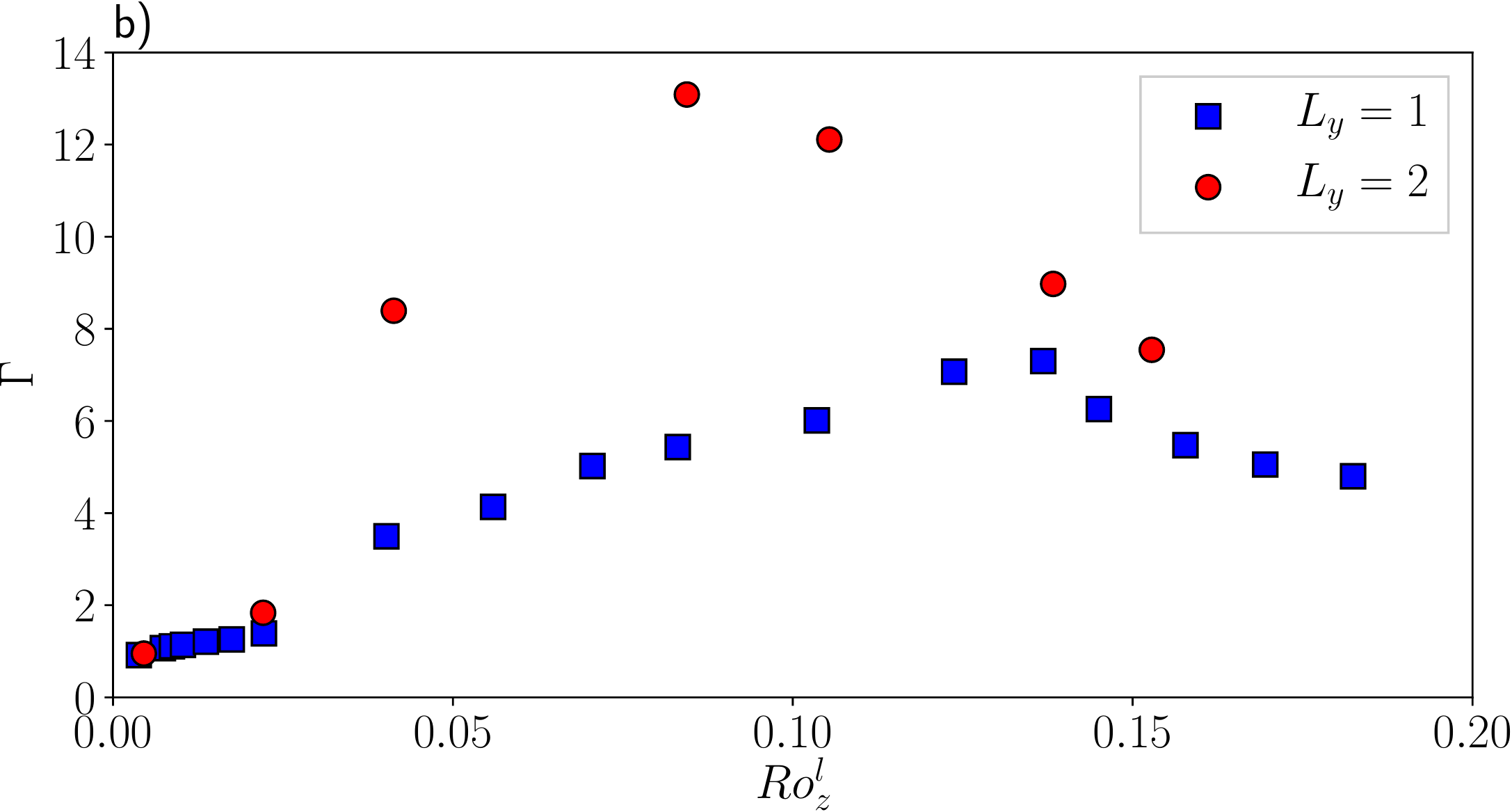}
       \caption{\label{fig:Gamma}
       $\Gamma$ as a function of (a) the normalised Rayleigh number and (b) the local Rossby number for $\Ek=10^{-5}$.}
\end{figure}

To test whether both of the conditions~(i) and (ii) apply to the formation of jets, 
we perform a number of simulations with increasing $\Ra$ for $\Ek=10^{-5}$
and with $(L_x,L_y)=(1,2)$ (listed in table~\ref{tab:existence}). Figure~\ref{fig:Gamma}a shows the value of $\Gamma$ as a function of $\Ra/\Ra_c$ for these cases,
which is compared with cases with $(L_x,L_y)=(1,1)$ at the same $\Ek$ as used in \citet{Guervilly2014}. 
$\Gamma$~increases well above unity for both $L_y=1$ and $L_y=2$ when $\Ra/\Ra_c>3$.
Condition~(ii) is therefore quantitatively the same for both jets and LSVs. Figure~\ref{fig:Gamma}b shows $\Gamma$ as a function of $\Ro_z^l$
and indicates that condition~(i) is quantitatively different for jets ($L_y=2$) and LSVs ($L_y=1$): the amplitude of the jets start to decay when the local Rossby number is greater than $0.08$,
which is smaller than the Rossby number at which $\Gamma$ starts to decline for the LSV. We thus conclude that the jets are more fragile than the LSVs when the local Rossby number approaches $0.1$.
Note that for the largest Rayleigh number computed for $L_y=2$, $\Gamma$ takes similar values for $L_y=1$ and $L_y=2$; however, the large-scale flow for $L_y=2$ still consists of an $n=1$ jet accompanied by LSVs in the mean cyclonic region.

\subsection{Effect on the heat transfer}
\label{sec:heat}

We anticipate that convective heat transfer might be impeded by the presence of jets due to either (i) the shear flow itself, or (ii) the local increase of the rotation rate in the regions of large-scale cyclonic vorticity. It should though be noted that the influence of a mean shear flow on convection is complex and may even enhance the efficiency of the heat transport \citep{CB_1992}. Effect~(i) is expected to be important when most of the convective heat transport is directed parallel to the mean velocity gradients. Such is the case outside the tangent cylinder 
in spherical geometry \citep[\eg][]{Yad16}, and in planar geometry when the rotation axis is perpendicular to the direction of gravity \citep[\eg][]{Hardenberg2015}. In our system, the effect of the shear is difficult to predict:
on the one hand, the convective heat flux is perpendicular to the mean velocity gradients, so the shear might not be particularly disruptive;
on the other, the elongation of the vertical flows in the $x$-direction observed in figure~\ref{fig:uzwz_Ly8}b indicates that the vertical velocity is affected by the shear. 
Effect (ii) is due to the well-known inhibiting effect of rotation on convection \citep{Chandrasekhar61} and has previously been
observed in the presence of LSVs when $L_x=L_y$; here the contribution of the LSV vorticity to the background rotation is not negligible locally \citep{Guervilly2014}.
The decrease in the heat transport efficiency associated with LSVs depends on the horizontal box size since wider horizontal domains 
promote the formation of broader LSVs of increasing strength.
In the case of jets, the efficiency of the heat transfer may also depend on their amplitude and lengthscale. 

\begin{figure}
\centering
       \includegraphics[clip=true,width=0.7\textwidth]{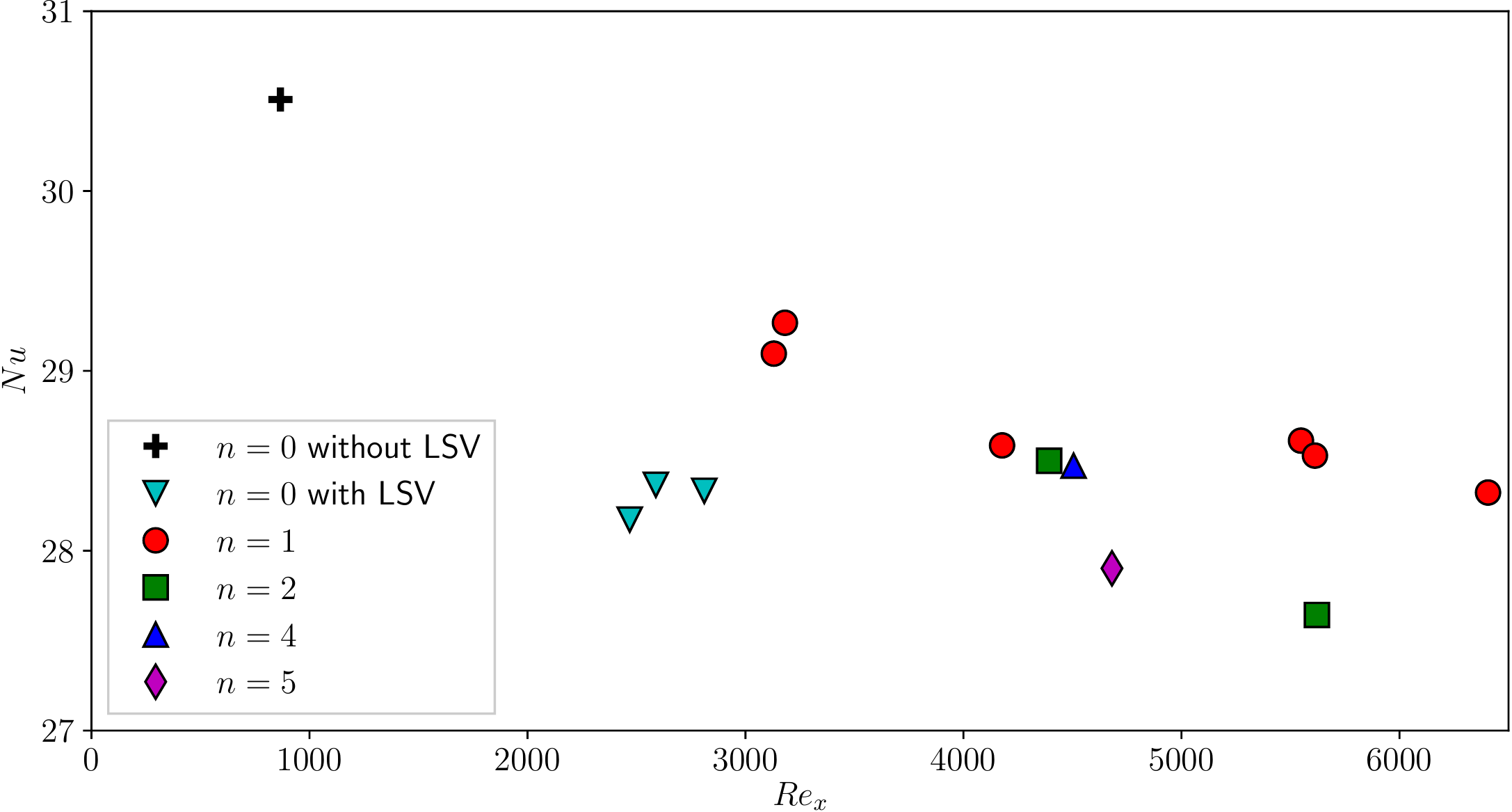}
       \caption{\label{fig:Nu}
       $\Nu$ as a function of the Reynolds number based on $u_x$ for the simulations of table~\ref{tab:Ek1e5} ($L_x=1$ and varying $L_y$) 
       and for a reference case without a large-scale flow ($L_x=L_y=0.3$).}
\end{figure}

The global efficiency of the heat transfer is often quantified by the Nusselt number, $\Nu$, which is a measure of the total heat flux through the 
layer normalised by the heat flux in the absence of convective motions. In our system, in the absence of heat sources and with fixed-temperature boundary
conditions, the Nusselt number is defined as 
\begin{equation}
	\Nu = 1 + \langle |\bnabla \theta |^2\rangle,
\end{equation} 
where the angle brackets denote time- and volume-averages.
To compare quantitatively the effect of the presence of large-scale flows on $\Nu$, we need to calculate $\Nu$ in a reference case without large-scale flows. 
This can be obtained by considering a small horizontal box size that accommodates only about three convective cells, thereby preventing the formation of a large-scale flow. 
For $\Ek=10^{-5}$ and $\Ra=3\times10^8$, the reference case is computed for $L_x=L_y=0.3$.
Figure~\ref{fig:Nu} shows the Nusselt number as a function of $\Rey_x$ for the simulations of table~\ref{tab:Ek1e5} and for the reference case.
Overall, $\Nu$ is smaller by approximately $10\%$ in cases with a large-scale flow than it is in the reference case.
As we might have expected, there is a decreasing trend of $\Nu$ with $\Rey_x$ for the cases with jets ($n>0$).  However, at similar values of $\Rey_x$, $\Nu$
is relatively unaffected by the number of jets. 
The cases with only an LSV ($n=0$) have smaller values of $\Nu$ than cases with jets with similar values of $\Rey_x$. 

\begin{figure}
\centering
       \includegraphics[clip=true,width=0.8\textwidth]{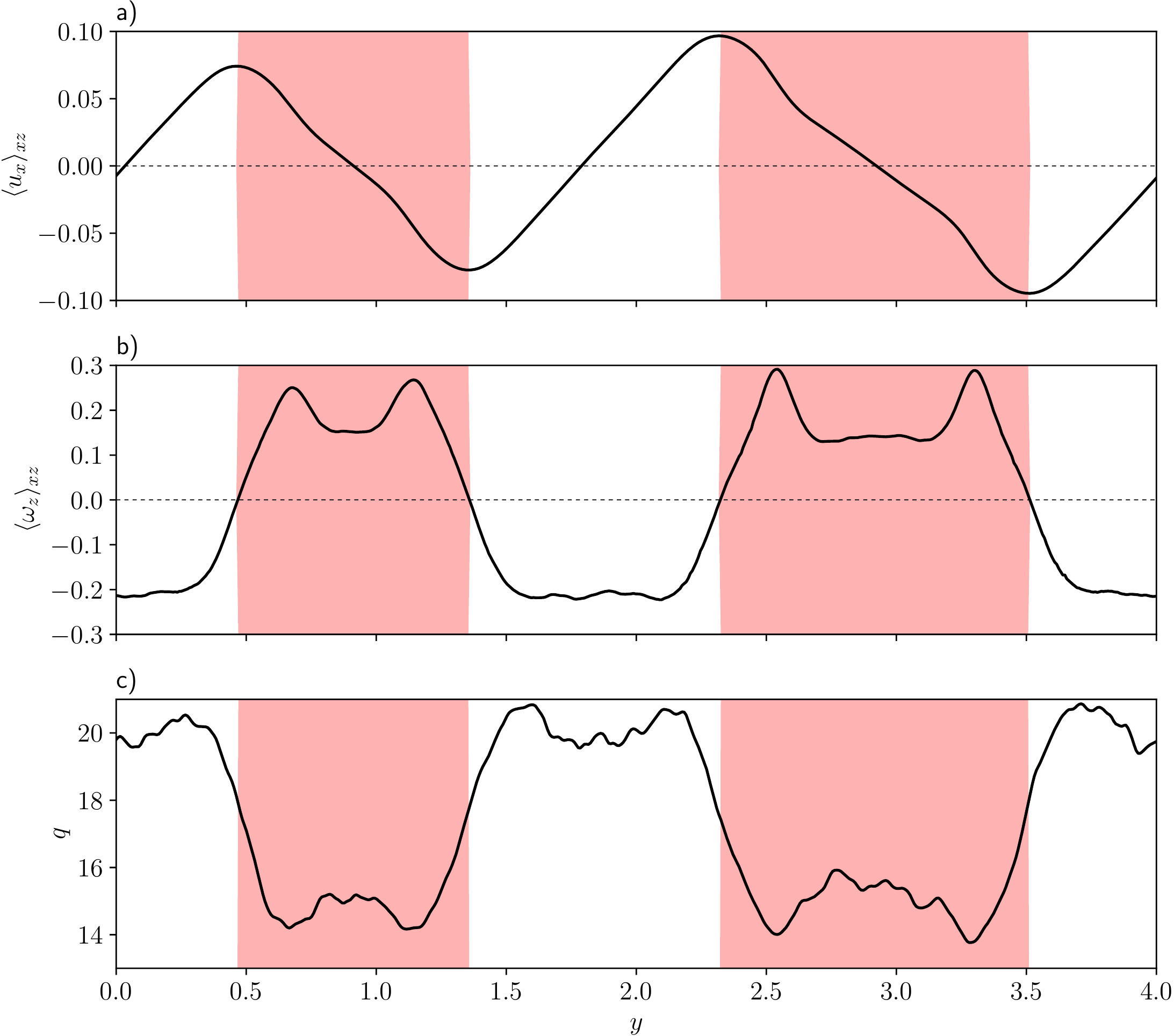}
       \caption{\label{fig:q_Ly4}
       Profiles of (a) the time-averaged mean flow, $\langle u_x \rangle_{xz}$, (b) the mean axial vorticity, $\langle \omega_z \rangle_{xz}$, and (c) the
       mean heat flux at the upper boundary, $q$; $\Ek=10^{-5}$, $\Ra=3\times10^8$, $L_x=1$ and $L_y=4$. The red bands correspond to 
       the regions of positive mean vorticity.}
\end{figure}

We now examine the local effect of the large-scale flows on the heat transfer. Figure~\ref{fig:q_Ly4} shows
the time-averaged heat flux at the upper boundary, $q=-\langle \partial_z T (z=1)\rangle_x$, for $(L_x,L_y)=(1,4)$.
For comparison, we also plot the time averages of $\langle u_x \rangle_{xz}$ and $\langle \omega_z \rangle_{xz}$.
The heat flux is lower by approximately $25\%$ in the regions of positive $\langle \omega_z \rangle_{xz}$ than in the regions of negative $\langle \omega_z \rangle_{xz}$.
This main difference is not due to the shearing of the convective flows by the jets because the highs of $q$
are not correlated with regions of smallest shear.
The heat flux is particularly low in the locations of the large-scale cyclone, which appear distinctly as peaks in the profile of $\langle \omega_z \rangle_{xz}$.
This detrimental effect of large-scale cyclonic flow on the heat transport is due to the local increase of rotation rate, as in the case of the LSVs when $L_x=L_y$. 
This is consistent with the fact that the mean axial vorticity persistently reaches values of $0.2$ (in units of $2\Omega$). 
In the reference case $L_x=L_y=0.3$, $q\approx18$ on average, so this argument also explains why the anticyclonic regions enhance the heat flux.
In cases with LSVs only ($L_y\in[1,1.05]$), the mean axial vorticity persistently reaches values of $0.5$, which explains why LSVs have a 
more pronounced effect than jets on $\Nu$ (figure~\ref{fig:Nu}).

\section{Conclusion}
\label{sec:ccl}

We have studied the formation of large-scale flows in rotating anisotropic convection; this has been achieved through numerical simulations in planar geometry, with horizontal anisotropy introduced by choosing a box aspect ratio with unequal horizontal dimensions ($L_y>L_x$).
For the standard case of $L_y/L_x=1$, which has received much attention in the literature \citep[\eg][]{Chan07,Kapyla11,Jul12,Favier2014,Guervilly2014,Rubio14,Stellmach2014}, 
the stable configuration consists of a depth-invariant large-scale vortex, formed by the clustering of small-scale convective vortices, and which grows to the horizontal box size.
When $L_y/L_x$ is increased above unity, the system undergoes a transition from the LSV-only configuration to one in which depth-invariant unidirectional flows emerge. 
As for LSVs, the necessary conditions for the formation of these unidirectional flows are small Rossby numbers (based on the convective lengthscale and velocity) and sufficiently large Reynolds numbers.
Over less than one global viscous timescale, persistent unidirectional flows with multiple jets are produced when $L_y \gtrsim 3 L_x$.
On longer timescales, the multiple jets merge to reach the largest available lengthscale in the system.
Interestingly, we found that LSVs of size comparable with $L_x$ systematically coexist with the jets. 
The LSVs are located in the flanks of the jets and can be persistent or intermittent depending on the lengthscale of the jet.
The transition from the LSV-only configuration to a configuration involving both jets and LSVs occurs for a small degree of anisotropy, namely $L_y/L_x\approx1.08$ when $\Ek=10^{-5}$ and $L_x=1$.
In the vicinity of this transition, the system is bistable; random changes between the two large-scale states are observed when the simulations are integrated 
over several viscous timescales.
This bistable regime is observed only for a narrow interval of the box aspect ratio ($L_y/L_x\in[1.06,1.15]$ for $\Ek=10^{-4}$ and $L_x=2$).

Large-scale flows are known to influence the efficiency of the heat transfer in rotating convection  \citep[\eg][]{Hardenberg2015,Yad16}. In our model, although there are no sizeable vertical flows associated with the presence of the LSVs or the jets, we find that the large-scale flows modify the transport properties of the convection by changing locally the rotation rate. As the axial vorticity of the large-scale flows is skewed towards cyclonic vorticity in our simulations, these flows reduce the efficiency of the convective heat transport. In more realistic planetary conditions at smaller Ekman and Rossby numbers, the skewness of the axial vorticity is expected to diminish, and hence the effect of the large-scale flows on the efficiency of the heat transfer may be different \citep{Jul12}.

\begin{figure}
\centering
       \includegraphics[clip=true,width=0.7\textwidth]{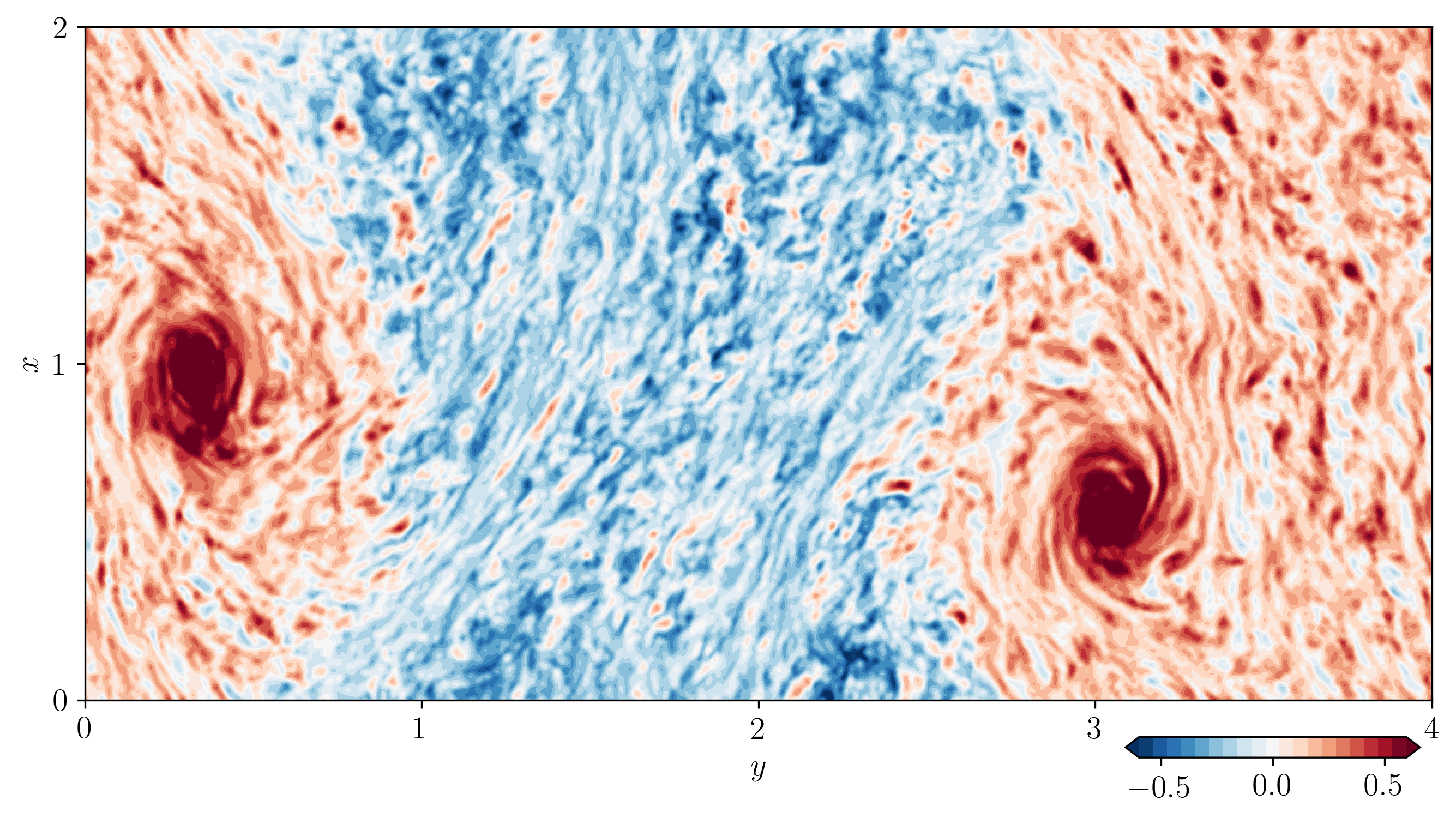}
       \caption{\label{fig:Lx2Ly4}
       Snapshot of the depth-averaged axial vorticity in the saturated regime for the case of $L_x=2$, $L_y=4$.
       }
\end{figure}

In all of the simulations described above, the domains are characterised by $L_x=1$, with differing values of $L_y$. Although a systematic exploration of $L_x $-$L_y$ space is not computationally feasible when both $L_x$ and $L_y$ exceed unity, it is nonetheless of interest to investigate the nature of the solution in the regime $1 < L_x < L_y$, given that, for $1 < L_x = L_y$, the attracting solution will consist of only an LSV. To this end, we simulated the case of $L_x=2$, $L_y=4$, with the standard parameter values of $\Ek=10^{-5}$, $\Ra=3\times10^8$ and $\Pran=1$. Following a small perturbation to the static state, the general form of the evolution is as described above for the cases with $L_x=1$, with an initial rapid growth and decline in the kinetic energy, followed by a slow evolution to the large-scale state (as shown, for example, in figure~\ref{fig:KE_Ly8}). In this extended domain, the slow evolution is accordingly lengthened, with equilibration of the kinetic energy not occurring until $t \approx 5 \times 10^4$. Figure~\ref{fig:Lx2Ly4} shows a snapshot of the depth-averaged axial vorticity in the long-time stationary regime, with one jet and with LSVs embedded in the flanks of the cyclonic component; the LSVs extend essentially across the entire $x$-domain. In comparison with the case of $L_x=1$, $L_y=2$, which has the same aspect ratio in the horizontal plane, the flows in the extended domain ($L_x=2$, $L_y=4$) are, as expected, much more vigorous, with, for example, $\Rey_x = 9870$, in comparison with $\Rey_x=5548$ in the smaller domain; as a result, this leads to a reduction in the heat transfer ($\Nu = 26.9$ in comparison with $\Nu=28.6$).

In spherical shell models of rotating convection in the molecular envelope of gas giant planets, shallow large-scale vortices have been observed to coexist with deep zonal flows (\ie flows that extend throughout most of the layer depth) \citep{Heimpel2016}. The persistence of deep LSVs when the horizontal symmetry is broken, together with their coexistence with jets as demonstrated here,
is a novel feature in rotating convection. 
We have used a simplified model that does not permit the saturation of the lengthscale of the large-scale flows as a result of either the curvature of the boundaries 
or the latitudinal variation of the rotation background. Our model is therefore unable to provide predictions for the amplitude and lengthscale of the large-scale flows in terms of the properties of the convective flows. However, it does offer a valuable proof of concept of the possible coexistence of deep LSVs and jets in rotating convection.

In numerical simulations using periodic boundaries, the dynamics should ideally not be influenced by the choice of the computational domain size. 
However, large-scale flows can be affected by confinement effects due to the finite domain size, as we find here. Numerical simulations in very
wide computational domains are required to study the influence of confinement on the large-scale dynamics, 
but, unfortunately, such simulations are currently out of reach of 3D models of rapidly rotating convection. 
The analogy between the large-scale dynamics observed in 3D simulations and in forced 2D turbulence simulations (\eg \citet{Frishman2017}) 
supports the study of the evolution of large-scale flows over long timescales using 2D turbulence models in the wide domains
that 3D models cannot explore. 
It would be particularly interesting to perform 2D simulations with a forcing possessing similar statistical properties to the small-scale convective buoyancy driving found in 3D models.  
An alternative and complementary approach to 3D direct numerical simulations,
proposed by \citet{Julien2007}, is to model rotating Rayleigh-B\'enard convection using an asymptotically reduced set of PDEs 
valid in the limit of small Rossby number. Certain results obtained with the reduced system are found to be in excellent agreement with 3D numerical simulations of the full equations run at moderately low
Ekman and Rossby numbers in the absence of domain horizontal anisotropy \citep[\eg][]{Jul12,Stellmach2014}. The reduced model, which assumes the dominance of the planetary vorticity over the planetary vorticity, is however unable to capture any asymmetry between cyclonic and anticyclonic motions. 
Interestingly, the effect of domain horizontal anisotropy on the formation of large-scale flows in rotating Rayleigh-B\'enard convection has been recently investigated 
with the reduced system by \citet{Julien2017}.
They find that unidirectional jets with embedded persistent LSVs form for aspect ratios $L_y/L_x$ in excess of $1.1$, in good agreement with our 3D numerical results.
The asymptotically reduced system of \citet{Julien2017} would therefore be a valuable tool to study the confinement effects of the large-scale dynamics 
in the limit of small Rossby number.

An alternative method of introducing horizontal anisotropy in a Cartesian domain is to tilt the rotation axis with respect to the direction of gravity.  This configuration would arguably offer a greater challenge for the formation of the LSVs because of the production of horizontal flows with a vertical shear 
due to the tilt of the convective rolls \citep{Hathaway1983}.
This configuration was investigated by \citet{Chan13} for compressible convection.
They observed the formation of LSVs in cases with both a small tilt of the rotation axis (corresponding to a latitude of $67.5^{\circ}$)  as well as a large tilt (a latitude of $22.5^{\circ}$).
The rotation rate required for the appearance of the LSVs is higher at low latitudes.
The role played by compressibility in the formation of the LSVs is difficult to assess in this previous work; it would therefore be of great interest to 
investigate the evolution of the LSVs in Boussinesq convection with a tilted rotation axis.

The collective action of LSVs and rotating convection can produce coherent large-scale magnetic fields at 
moderate magnetic Reynolds numbers and low Prandtl numbers (the parameter regime relevant for planetary dynamos)
despite the disruptive feedback of magnetic fields on coherent Reynolds stresses \citep{Guervilly2015, Guervilly2017}.
This magnetic feedback is known to strongly affect unidirectional flows \citep[\eg][]{Aub05,Tobias07}.
It would therefore be of considerable interest to investigate whether the jets can promote the dynamo mechanism driven by the LSVs and the convection
or whether instead they are completely suppressed by the self-sustained magnetic fields. 
We shall pursue this question in a forthcoming study. 

\section*{Acknowledgements}
CG was supported by the Natural Environment Research Council under grant NE/M017893/1. Computations were performed using the ARCHER UK National Supercomputing Service (\url{http://www.archer.ac.uk}); the facilities of N8 HPC Centre of Excellence, provided and funded by the N8 consortium and EPSRC (Grant EP/K000225/1) and co-ordinated by the Universities of Leeds and Manchester; and ARC1, part of the High Performance Computing facilities at the University of Leeds.
We are grateful to Keith Julien and co-authors for sharing with us the preprint of their recent paper. 
We thank the two anonymous referees for suggestions that have improved the manuscript.


\end{document}